\documentclass[useAMS,usenatbib,subeqn]{mn2e}

\usepackage{multirow}  
\usepackage{color}
\usepackage{epsfig,amssymb}
\usepackage{graphicx,epsfig}
\usepackage{hhline}
\usepackage{subfigure}
\usepackage{eufrak}
\usepackage{mathrsfs}
\usepackage{eucal}
\usepackage{amsmath}
\usepackage{aas_macros}
\usepackage{fontenc}
\usepackage{textcomp}
\usepackage{natbib}
\usepackage{hyperref}

\title{The VLT Survey Telescope ATLAS} \author[T. Shanks et al.]
{T. Shanks$^{1}$\thanks{E-mail: tom.shanks@durham.ac.uk}, N. Metcalfe$^1$,
B. Chehade$^{1}$,  J.R. Findlay$^{1}$, M.J. Irwin$^2$,
\newauthor E. Gonzalez-Solares$^2$, J.R. Lewis$^2$, A. Kupcu Yoldas$^2$,
\newauthor R.G.Mann$^3$, M.A. Read$^3$, E.T.W. Sutorius$^3$, S. Voutsinas$^3$ \\ 
$^{1}$Physics Department, University of Durham, South Road, Durham, DH1 3LE, UK \\
$^{2}$Institute of Astronomy, Univ. of Cambridge, Madingley Road, Cambridge, CB3 0HA, UK \\
$^{3}$Institute for Astronomy, Univ. of Edinburgh, Blackford Hill, Edinburgh EH9 3HJ, UK.
}
\begin{document}

\pagerange{\pageref{firstpage}--\pageref{lastpage}} %\pubyear{2014}

\maketitle 
\begin{abstract}

The VLT Survey Telescope (VST) ATLAS  is an optical $ugriz$ survey
aiming to cover $\approx4700$deg$^2$ of the Southern sky to similar
depths as the Sloan Digital Sky Survey (SDSS). From   reduced images and
object catalogues provided by the Cambridge Astronomical Surveys Unit we
first find that the  median seeing ranges from $0.''8$ FWHM in $i$  to
$1.''0$ in $u$, significantly better than the $1.''2-1.''5$ seeing for
SDSS. The 5$\sigma$ magnitude limit for stellar sources is $r_{AB}=22.7$
and in all bands these limits are at least as faint as SDSS. SDSS and
ATLAS are more equivalent for galaxy photometry except in the $z$ band
where ATLAS has significantly higher throughput. We have improved the
original ESO magnitude zeropoints by comparing $m<16$ star magnitudes
with APASS in $gri$, also extrapolating into  $u$ and $z$, resulting in
zeropoints accurate to $\approx\pm 0.02$mag. We finally compare star and
galaxy number counts in a 250deg$^2$ area with SDSS and other count data
and find good agreement. ATLAS data products can be retrieved from the
ESO Science Archive, while support for survey science analyses is
provided by the OmegaCAM Science Archive (OSA), operated by the
Wide-Field Astronomy Unit in Edinburgh.
\end{abstract}


%=============================================================%
\section{Introduction} 
%=============================================================%

The ATLAS survey \citep{shanks13} is carried out on the VLT Survey
Telescope (VST), a 2.61-m telescope situated at Cerro Paranal
Observatory \citep{schipani12}. It uses the OmegaCAM camera\citep
{kuijken11} containing 32 4k$\times$2k CCDs with a total of 268
megapixels. The field-of-view of the camera is $1\times1$deg$^2$ and the
pixel size is 0.21arcsec. The aim of the survey is to produce a $ugriz$
catalogue to the equivalent depth of SDSS but in the Southern
Hemisphere, with a target area of $\approx4700$deg$^2$. Although the
survey was only specified to have seeing in the range  $1-1.''4$ FWHM,
with the better seeing going to the VST KiDS survey \citep{dejong13},
ATLAS is proving to have better median seeing than expected, at the
subarcsecond level.  The survey is being made alongside its sister VISTA
Hemisphere Survey (VHS, \citealt{mcmahon13}) which is supplying YJK.
The survey depths are also well matched to the WISE survey
\citep{wright10} at the mid-IR L and M bands. The footprint of the
survey is shown in Fig. \ref{fig:footprint} where it is compared to the
footprints of KiDS and VHS. The deeper VISTA VIKING NIR survey
\citep{edge13} has approximately the same footprint as KiDS. ATLAS has yearly
data releases with DR1 occurring in April 2013 and covering $\approx1500$deg$^2$ 
of data  taken before 30/9/12 and DR2 now imminent and covering 
$\approx2500$deg$^2$ taken before 30/9/13.

%=============================================================%
\subsection{ATLAS Science Aims}
%=============================================================%
The primary aims of the ATLAS survey are cosmological. The UV
sensitivity of the survey gives it an advantage in terms of quasar
surveys that in  the Southern Hemisphere can be followed up using
instruments like AAT 2dF or VISTA 4MOST. These surveys can then be
used, for example, to look for any evidence of excess power in the
quasar clustering data that might provide evidence for primordial
non-Gaussianity in the early Universe. It may also be possible to search
for BAO at the 100Mpc scale in the unexplored $0.8<z<2.2$ redshift
range.  These analyses can be done either by photo-z using  selection
codes like XDQSO-Z \citep{bovy12} or by using spectroscopic redshifts. 
We have already used AAT 2dF to make a redshift survey of $\approx10000$
quasars using UVX and NIR selection and these are  being used to study
the luminosity and clustering correlation function of dusty and
unobscured quasars (Chehade et al 2014 in prep.). This quasar redshift
survey can also form the base for photo-z clustering analyses over the
whole VST ATLAS area. Such quasar surveys will also complement future 
programmes such as the X-ray AGN surveys of e-ROSITA \citep{merloni12}.

Another cosmological aim is to test for the ISW effect by cross-correlating
the 2-D positions of Luminous Red Galaxies with  microwave background
temperature fluctuations e.g. \citep{sawangwit10}. This is one of the few
independent tests of the accelerating expansion of the Universe and the
Southern Hemisphere is the only place left to improve the statistical
significance of this test.

The galaxy and quasar surveys can also be combined to study the  quasar
lensing magnification bias at high redshift and the quasar galaxy
clustering environment at low redshifts (e.g. \citealt{mountrichas07}). 
ATLAS will also take advantage of its excellent seeing to allow many
galaxy lensing shear projects. Although not as deep as KiDS, ATLAS
shares the same VST platform which means that the advantages of low
optical distortion are shared by ATLAS for lensing projects.

We shall also be looking at galaxy counts and how they vary over the sky
at bright magnitudes. Previous observations from 2MASS have suggested
that there exists a galaxy  under-density in the Southern Hemisphere
\citep{keenan13,whitbourn14b} and we wish to test for the existence of
this `Local Hole' at optical wavelengths. Since VST ATLAS will also
cover the Great Attractor region, this means that it can use the form of
bright galaxy counts to search a large area in this direction for
clusters and superclusters even behind Shapley 8.

Mapping the Southern Hemisphere using galaxy clusters is also another
cosmological aim of VST ATLAS. The $riz$ bands will be combined with VHS
and WISE NIR bands to select galaxy clusters to $z\approx1$.

There are many other  non-cosmological projects that can be done with
ATLAS. The discovery of dwarf satellite galaxies and stellar streams in
the Southern Hemisphere is one clear example \citep{belokurov14,
koposov14} and ATLAS is also being used to search for high redshift
quasars via the Lyman-break technique (Carnall et al, 2014a, in prep.)

%=============================================================%
\section{ATLAS Description}
%=============================================================%
\subsection{Survey area }

\begin{figure}
\centering
%\resizebox{\hsize}{!}{\includegraphics{vhstest.png}}
\resizebox{\hsize}{!}{\includegraphics{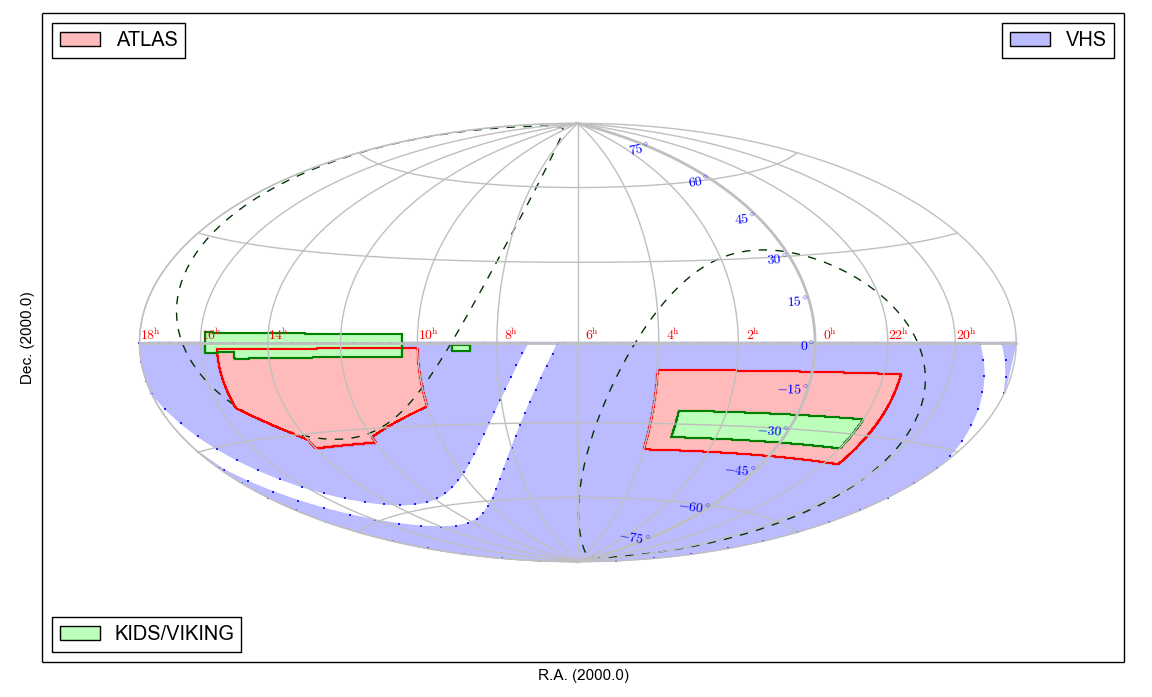}}
\caption{ATLAS footprint (red) compared to VHS (purple) and KIDS(green). VHS surveys the 
whole hemisphere except at $\pm 5$ deg. from the Galactic plane.}
\label{fig:footprint}
\end{figure}

As can be seen from Fig. \ref{fig:footprint},  the ATLAS South Galactic Cap
area (SGC) area lies between 21h30 $< $RA $<$ 04h00 and $-40<$ Dec $<
-10$. The NGC area lies  between 10h00$ < $RA $< $15h30 and $-20 <$ Dec
$ < -2.5$ plus 10h00 $<$ RA $<$ 15h00 and -30 $< $Dec $<$ -20. The total
area of the survey is 4711 deg$^2$ with 2087 deg$^2$ in the NGC and 2624
deg$^2$ in the SGC. The NGC area below Dec$<$-20 is so far approved in
$iz$ and is the subject of a Chilean ESO proposal in $ugr$. There is also
an ongoing `Chilean $u$ extension project' to double the  exposure time in
$u$  from 2 mins to 4 mins over the full ATLAS area (PI L. Infante).
Status maps of the ATLAS survey can be found at \url{http://astro.dur.ac.uk/Cosmology/vstatlas/}.

\begin{figure}
\centering
%\resizebox{\hsize}{!}{\includegraphics{vhstest.png}}
\resizebox{\hsize}{!}{\includegraphics{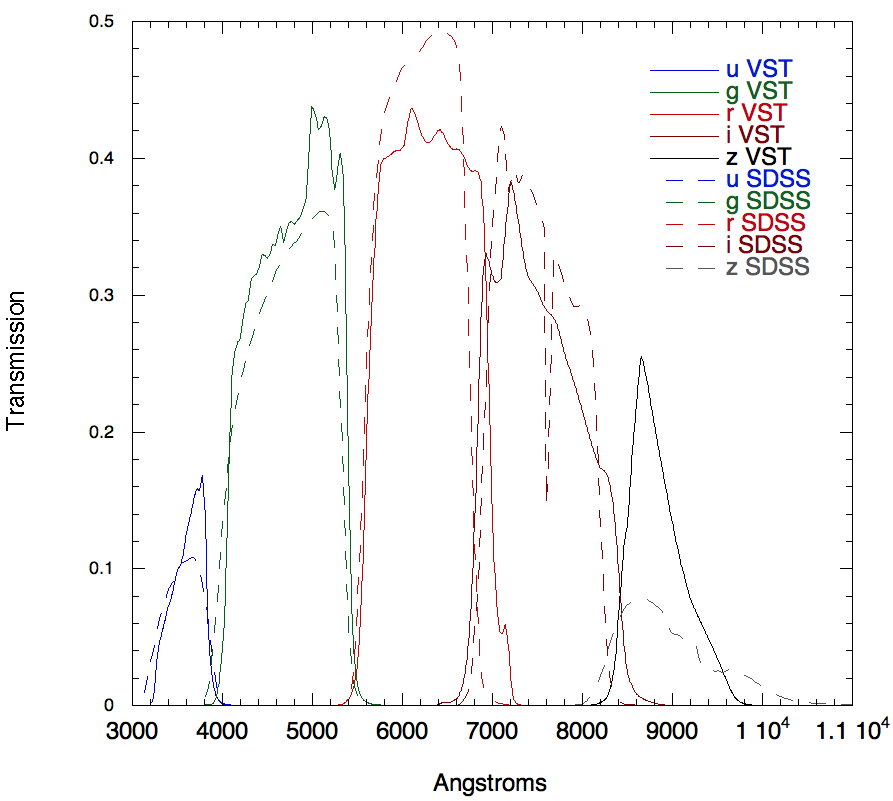}}
\caption{ATLAS and SDSS filter $+$ telescope $+$ atmospheric  transmissions compared. All are calculated at airmass 1.3. 
No Atmospheric Dispersion Corrector is assumed for the ATLAS filters since it is not being used.}
\label{fig:filters}
\end{figure}

\subsection{Survey observations}
\begin{figure}
\centering
%\resizebox{\hsize}{!}{\includegraphics{conf_v2.jpg}}
\resizebox{\hsize}{!}{\includegraphics{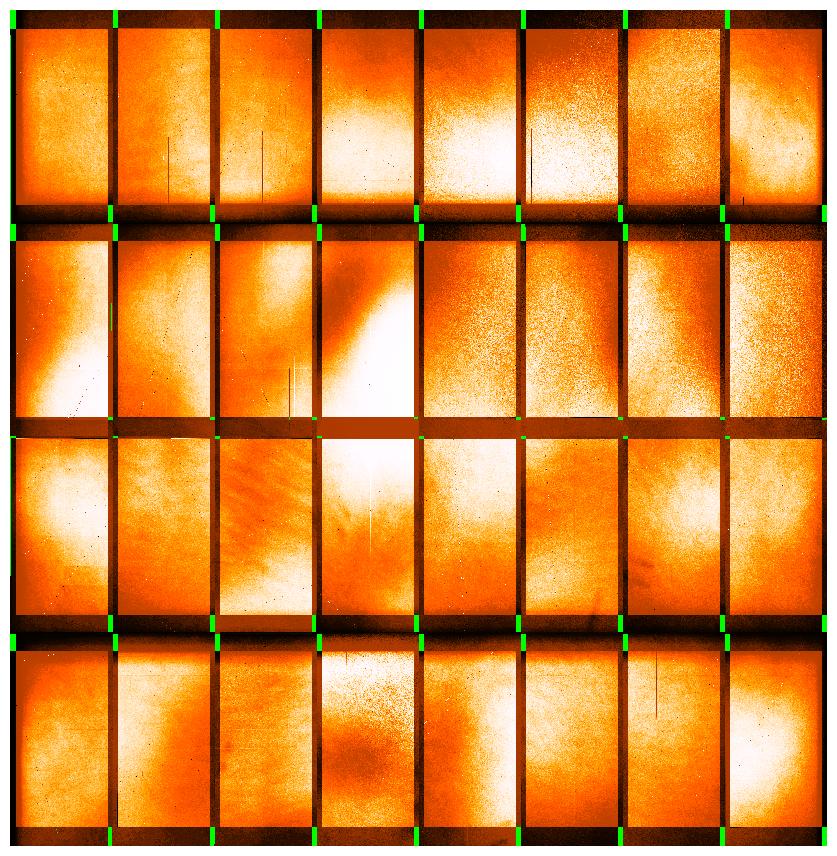}}
\caption{Typical ATLAS confidence map for a stacked tile in $r$. Areas left 
uncovered by the 2-pawprint dither are shown in green. Those at the perimeter
of the tile area will be covered by the $2'$ tile overlaps.}
\label{fig:conf}
\end{figure}

OmegaCAM  camera pixels are $0.213\times0.213$arcsec$^2$ in size so
approximately half the dimensions of  $0.396\times0.396$arcsec$^2$ SDSS
pixels. The $ugriz$ band transmissions are shown in Fig.
\ref{fig:filters} where they are compared to those for SDSS. They are
seen to be similar in all bands except for $z$ where ATLAS has an
$\approx2\times$ higher throughput. The ATLAS exposure times have  been
conservatively increased over the typical SDSS 54s exposure to maintain
the S/N achieved by SDSS  taking into account increased read-out noise
and, potentially, sky brightness, particularly in $i$ and $z$ (see Table
\ref{tb:exp_time}). The OmegaCAM read-out noise  is typically 5-6
electrons per pixel and the gain is $\approx2.2$ with 10\% chip-to-chip
RMS. Double exposures are taken, dithered by $85''$ in Declination and
$25''$ in RA, to cover the main inter-chip gaps ($80.''5$ and $11.''8$
in Declination and $21.''5$ in RA)  and to allow cosmic rays to be
rejected. The 2-pointing dither  and $2'$ tile overlaps leaves 28 ($80.''5\times21.''5$)
and 14 ($11.''8\times21.''5$) small holes (see green areas in Fig. \ref {fig:conf}) amounting to
$\approx1/3$\% of the total area.    Each OmegaCAM field is
$\approx1\times1$deg$^2$ and each tile has a $58'$ centre-to-centre
spacing, giving a $\approx2'$ overlap between tiles in both the RA and
Dec directions. 

For each VST ATLAS tile, 2 dithered exposures in each of the $ugr$ bands
were observed in dark time and 2 dithered exposures in each of the $iz$
bands in grey/bright time. For all except the first 2 months of the survey, 
the observations were done in concatenations of 17 fields which approximately 
filled a 1hr ESO Observing Block (OB) in $u$ (including overheads) and slightly   less in the other bands.
The seeing is specified to be $<1.''4$ FWHM.

\begin{table}
\centering
\begin{tabular}{cccccc}
\hline\hline
Band & $u$ & $g$ & $r$ & $i$ & $z$ \\
\hline\hline

Exposure       & $2\times60$s & $2\times50$s & $2\times45$s & $2\times45$s & $2\times45$s \\
Seeing         & $1.''02$ & $0.''95$ & $0.''90$ & $0.''81$ & $0.''84$ \\
%SDSS Seeing   & $1.''63$ & $1.''52$ & $1.''43$ & $1.''37$ & $1.''40$ \\ From DR1
SDSS Seeing    & $1.''46$ & $1.''36$ & $1.''24$ & $1.''18$ & $1.''20$ \\ %Bramich & Freudling
%Mag Lim       & $22.10$ & $23.14$ & $22.71$ & $22.03$ & $20.85$ \\ %AB-Vega=1.0, -0.1, 0.2, 0.4, 0.5 
Mag Lim        & $21.99$ & $23.14$ & $22.67$ & $21.99$ & $20.87$ \\ %empirical median in r=1arcsec 
%Mag Lim (star)& $22.18$ & $23.04$ & $22.40$ & $21.58$ & $20.42$ \\% calculated in r=1arcsec assuming 1"seeing apcor
%Mag Lim (gal) & $21.66$ & $22.59$ & $22.02$ & $21.25$ & $20.80$ \\% calculated r=SDSS seeing (old)
Mag Lim (gal)  & $21.78$ & $22.71$ & $22.17$ & $21.40$ & $20.23$ \\% calculated r=SDSS seeing
%SDSS Mag Lim  & $21.00$ & $22.30$ & $22.00$ & $20.90$ & $20.00$ \\
%SDSS Mag Lim  & $21.95$ & $23.02$ & $22.75$ & $21.88$ & $20.45$ \\ This is for 1.4arcsec seeing
%SDSS Mag Lim  & $21.77$ & $22.85$ & $22.71$ & $21.95$ & $20.43$ \\ % Scaled by seeing 1.63-1.40
%SDSS Mag Lim   & $21.75$ & $22.63$ & $22.15$ & $21.62$ & $20.00$\\ % From count/s in seeing radius at am=1.0
SDSS Mag Lim   & $21.87$ & $22.75$ & $22.31$ & $21.71$ & $20.17$ \\ %Bramich and Freudling
%Sky Brightness& $21.56$ & $21.90$ & $20.96$ & $19.82$ & $18.83$ \\ AB-Vega=1.0, -0.1, 0.2, 0.4, 0.5
Sky Brightness & $22.34$ & $21.90$ & $20.92$ & $19.78$ & $18.85$ \\
SDSS Sky Bri.  & $22.15$ & $21.85$ & $20.86$ & $20.20$ & $19.00$ \\
$20^m$ $e^-/s$ & $29$    & $177$   & $160$   & $101$   & $29$ \\
SDSS $20^m$ $e^-/s$& $33$& $175$   & $174$   & $116$   & $19$ \\
\hline
%\hline\hline
\end{tabular}
\caption[]{ATLAS basic characteristics. ATLAS median seeing for ESO A,B
classified tiles. SDSS median seeing is taken  from \cite{bramich12}.
Mag Lim (ATLAS) corresponds to the median $5\sigma$  magnitude detection
limit for stars as measured in a $1''$ radius aperture. Mag Lim (gal)
corresponds to ATLAS 5$\sigma$ limit now calculated in apertures of the
SDSS seeing radius at airmass 1.0. SDSS Mag Lim corresponds to $5\sigma$
SDSS point source detection limits  based on the above fluxes and
calculated in a radius of the above median SDSS seeing at airmass 1.0.
Sky brightness is measured in mag/arcsec$^2$. SDSS DR1 sky brightness is
median of 77068 frames from
\url{http://www.sdss2.org/dr7/products/general/seeing.html}. Fluxes are
given at $20^m$ for ATLAS and SDSS normalised to airmass 1.3. SDSS
fluxes come from \cite{stoughton02} eqn. (1). All magnitudes are 
quoted in the AB system.
}
 \label{tb:exp_time}
\end{table}
%\url{http://www.sdss2.org/dr7/products/general/seeing.html}. 

The imaging data has been reduced by CASU using the  VST data flow
software. The images were trimmed and debiased using nightly calibration
frames and then flat-fielded using accumulated monthly stacked twilight
sky flats. The frames are then corrected for crosstalk and defringed if
necessary. The sub-exposures are then automatically registered and
stacked. The resulting imaging data comprises the combination of the two
individual images for each of the original CCDs. Each file is in a
multi-extension fits  (MEF) format with an extension for each of the 32
OmegaCam CCDs in the stacked tile. Individual CCDs originally contained
$2048\times4096$ pixels and the stacked pawprint extensions contain 
$\approx2165\times4498$ pixels to cover the two $25''\times85''$ offset
CCDs that make up each extension in a stack. This leaves $\approx5''$
overlaps between the stacked pawprints in each direction where objects
can be recorded twice. Along with the imaging data, statistical
confidence maps in the same format are also supplied \citep{irwin04}. In
Fig. \ref{fig:conf} we show a typical confidence map from the $r$ band.
The main `bar-like' pattern seen is due to the interchip gaps and the
dither pattern used to cover them. Note that these are detector-level
normalised confidence maps whereby each detector's map is normalised to a
median level of 100\%, hence the particular shape of the repeating
pattern.

Object detection is then carried out  to an isophotal limit set to
$1.25\sigma$ where $\sigma$ is a robust (rms) estimate of the
average pixel noise over the frame. Catalogues are then produced for the
stacked and unstacked images. Aperture, Petrosian and Kron magnitudes
are supplied in the first instance along with many other parameters (see
Section \ref{sect:params} below).

\subsection{Astrometry}
\begin{figure}
\centering
\resizebox{\hsize}{!}{\includegraphics{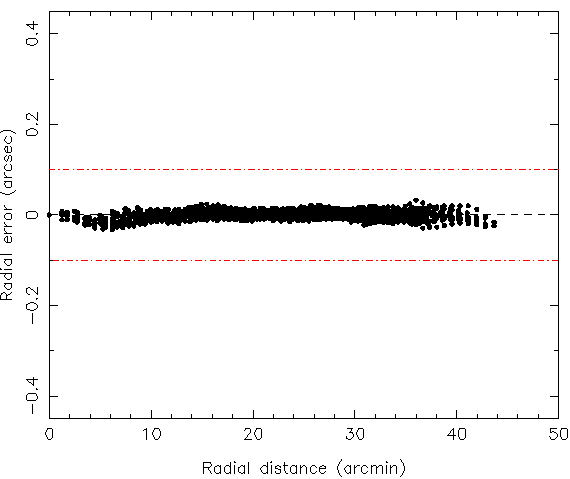}}
\caption{ATLAS astrometric accuracy from 2MASS comparison. The red dashed lines represent
reference levels of $\pm100$mas astrometric accuracy. }
\label{fig:astrometry}
\end{figure}

Astrometric calibration is via the numerous unsaturated 2MASS point
sources available in each field. By stacking residuals from a series of
standard Tangent Plane astrometric fits based on 2MASS we can see (as in
the example in Fig. \ref{fig:astrometry}) that there are no significant astrometric
distortions over the whole field of view.   The individual detector 
astrometric solutions achieve rms accuracies of around 70-80mas per star
- generally dominated by rms errors in 2MASS stars. Even at high
Galactic latitudes there are sufficient calibrators to give systematic
residuals at the ~25mas level per detector.  The global systematics from
stacking multiple solutions are better than this as can be seen in Fig.
\ref{fig:astrometry}.   A Tangent Plane projection (TAN) is being used for all data
products.

\subsection{Illumination correction} 
\begin{figure}
\centering
%\resizebox{\hsize}{!}{\includegraphics{i_illumn_correction.png}}
\resizebox{\hsize}{!}{\includegraphics{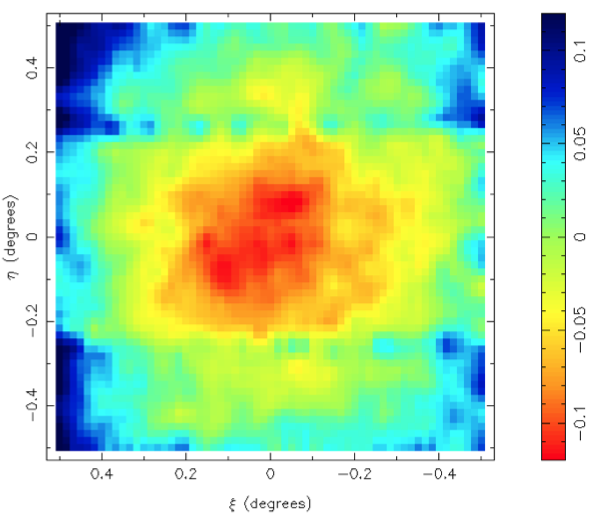}}
\caption{ATLAS illumination correction.  ATLAS $i$-band observations
processed using the same master flats for the period 16/8-30/9/12 and
the resulting  residuals from  the 2MASS photometric catalogue
stacked over several hundred independent pointings. The $i$-band colour
equations are $i = J + (J-K)*1.175 + 0.459$ (for $H-K \le 0.15$) and $i
= J + (J-K)*1.175 + 0.459 + (H-K-0.15)*3.529$   (for $H-K >  0.15$).  
  }
\label{fig:illum}
\end{figure}

The open nature of the VST dome
provides excellent through-flow of air improving the seeing by reducing
the contribution from the dome. However, this also leads to increased
scattered light which impairs flat-fielding. In particular a
$\approx0.2$mag centre to the edge gradient in the photometry is seen
from the pawprint. An illumination correction therefore  has to be
applied to the photometry and this has been done  via the AAVSO
Photometric All-Sky Survey (APASS)  survey
(\url{http://www.aavso.org/apass}). The APASS  survey  is a $g<16$mag
stellar survey made in the $BVgri$ bands. The illumination corrections
are obtained from direct, stacked, comparisons between ATLAS and APASS stelar magnitudes for $gri$
and from extrapolating APASS $gri$ to provide the $u$ and $z$
corrections. The pattern of residuals across  the tile typically looks
like that shown in Fig. \ref{fig:illum}. The scattered light is made up
of multiple components having different symmetries and scales causing
effects ranging in scale from ten arcseconds  with $x-y$ rectangular
symmetry, e.g. due to scattering off masking strips of CCDs, to large
fractions of the field due to radial concentration in the optics and to
non-astronomical scattered light entering obliquely in flats. The
illumination correction removes the dominant reproducible components of
this effect in the source lists leaving the zeropoint across the field
uniform to $\approx\pm0.007$mag. The illumination correction is updated
on a few months timescale as necessary.
 
\begin{figure}
\centering
%\resizebox{\hsize}{!}{\includegraphics{i_sdss_10stack_v2.png}}
\resizebox{\hsize}{!}{\includegraphics{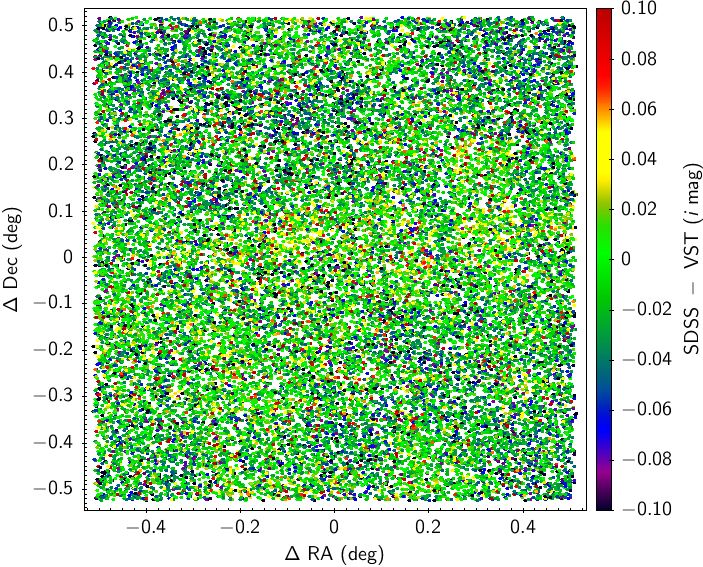}}
\caption{SDSS-ATLAS stellar residuals in $i$ from a  10-field stack
after ATLAS illumination correction.}
\label{fig:illum_fixed}
\end{figure}

We can test the illumination corrections in the overlap area with SDSS. 
Fig. \ref{fig:illum_fixed} shows SDSS-ATLAS residuals from a
 stack of 10 ATLAS $i$-band tiles in the overlap area. The amplitude of the 
radial pattern of  residuals is much reduced from that shown in 
Fig. \ref{fig:illum}.

\subsection{ATLAS zeropoint  calibration}
\label{sect:calibn}

The original zeropoint calibration was based on the ESO nightly
standards in all bands and these were used to place the VST magnitudes for each tile 
on a Vega-like system. These standards are observed in any photometric
conditions and so  cloud can introduce zeropoint error. However, due to
the excellence of the Paranal site and the clear conditions specified
for ATLAS observations, there appears to be a reasonable consistency
between the zeropoints in night-to-night comparisons. So these ESO
standard zeropoints make a good first-order calibration of the survey.
But in making stellar colour-colour diagrams Chehade et al (2014 in
prep.) found that the stellar locus moved on a regular basis, presumably
due to non-photometric conditions in one of the bands between survey
field and standard star. Therefore there is a need for an improved
global calibration for ATLAS and we have thus  made a first iteration
towards a global calibration via APASS.

We  used APASS to produce new magnitude zeropoints based on a 
tile-by-tile comparison of unsaturated $g<16$ mag stars in each ATLAS
$gri$ tile. In the $u$ and $z$ bands where there is no APASS data, we
again extrapolated from the APASS $gri$ bands band  to produce the $u$ and $z$
calibrations. In APASS, the average star sky density is
$\approx100$deg$^{-2}$ although at high latitudes, the star density may
be lower. Therefore to improve the statistics we have also implemented a
nightly ATLAS zeropoint based on all  the APASS standards observed that
night. This also has the effect of averaging out any APASS systematic 
in an individual ATLAS field. We now assume the nightly zeropoint as our
default calibration. This calibration can be checked in the SDSS
sub-areas and we will report on the results of this check in Section
\ref{sect:atlas_sdss}.

Note that despite the fact that the original ESO zeropoints are in the
Vega magnitude system, APASS uses the AB system and so the default for
this  paper is that the magnitudes are on the APASS AB system unless
explicitly stated otherwise. The AB-Vega magnitude offsets were
computed to be 0.894 ($u$), -0.100 ($g$), 0.159 ($r$), 0.356 ($i$), 0.517
($z$). These were calculated for VST telescope and filter throughput and
at airmass 1.3 (see Fig. \ref{fig:filters}) for consistency with SDSS conversions.

\begin{figure}
\centering
%\resizebox{\hsize}{!}{\includegraphics{apass_offset_hist.jpg}}
%\resizebox{\hsize}{!}{\includegraphics{APASS_ESO_offset_reduced_range.png}}
\resizebox{\hsize}{!}{\includegraphics{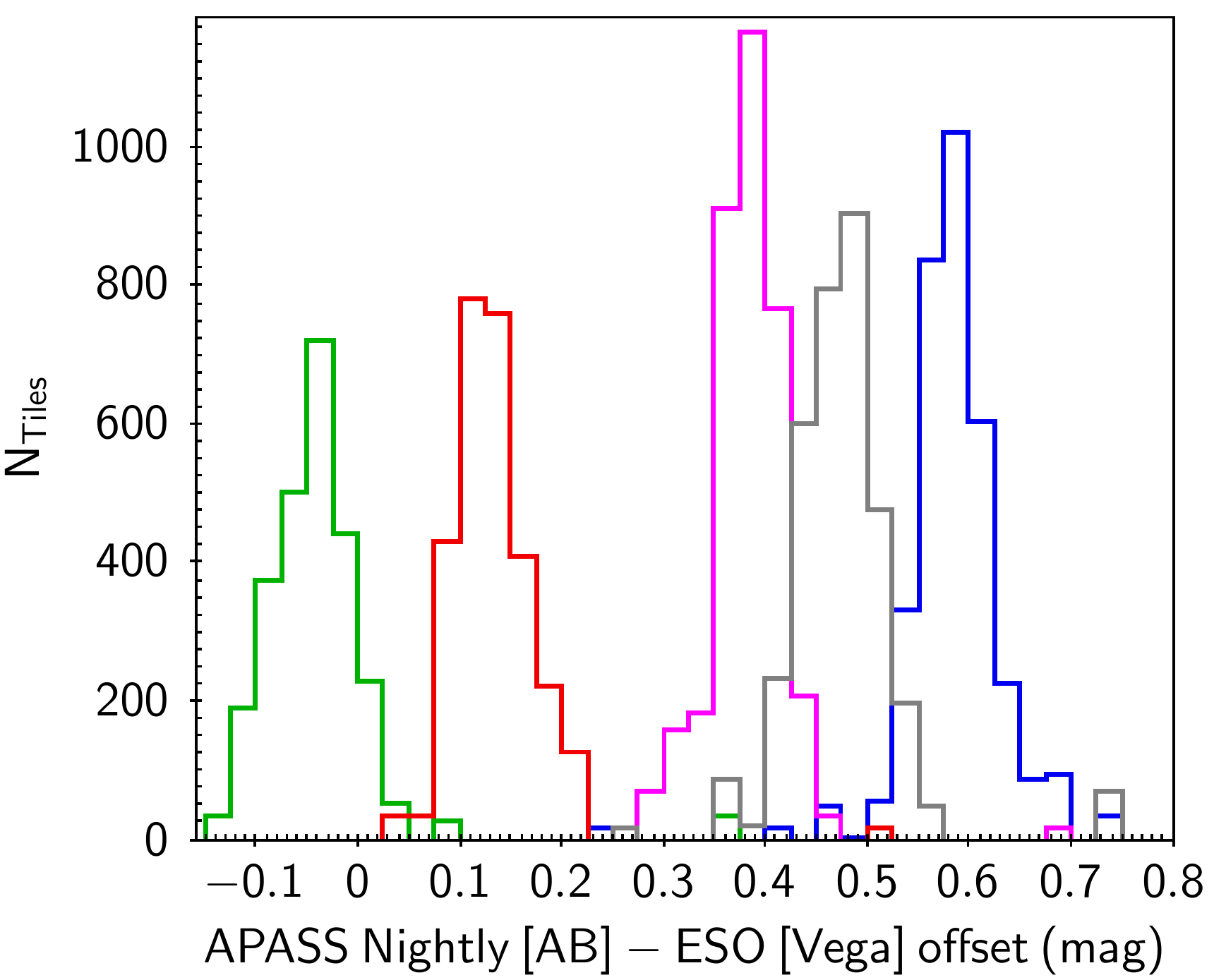}}
\caption{APASS nightly-ESO ATLAS magnitude zeropoint offsets on a
tile-by-tile basis over the whole current survey for  $u$ (blue), $g$ (green), $r$ (red), $i$
(purple), $z$ (grey). Here APASS is in AB whereas ESO ATLAS is in the
Vega system. The average APASS-ESO magnitude offsets and standard
deviations are $0.59\pm0.06$ ($u$), $-0.042\pm0.06$ ($g$),
$0.13\pm0.045$ ($r$), $0.39\pm 0.038$ ($i$), $0.47\pm 0.080$ ($z$).}
\label{fig:apass_hist}
\end{figure}

We then compared APASS nightly  magnitude zeropoints with the original
ESO zeropoints. Fig \ref{fig:apass_hist} shows distributions of
APASS-ATLAS magnitude zeropoint differences on a field-by-field basis
over the whole current survey area. We find that the residuals are
basically at the $\pm0.05$mag level but with non-Gaussian tails usually
arising from  non-photometric conditions. From Fig. \ref{fig:apass_hist}
we also note that the APASS-ESO offsets are in good agreement with the
expected offsets except in the case of $u$ which has a 0.3mag
discrepancy. It is probable that the $u$ band fabricated from APASS
$gri$ contains a zeropoint error and this is confirmed in the comparison
with SDSS made in Section \ref{sect:SDSS_col_equ} below. This problem
will be corrected at the global calibration stage (Findlay et al 2015,
in prep).

\subsection{ATLAS photometric parameters}
\label{sect:params}

The photometric quantities supplied for ATLAS are Kron and Petrosian
pseudo-total magnitudes along with their respective radii. Kron and Petrosian 
magnitudes are measured in these radii multiplied by a factor of $2\times$.
Aperture magnitudes are also given with radii of 1/2$\times r_{core}$,
1/sqrt(2)$\times r_{core}$, $r_{core}$, sqrt(2)$\times r_{core}$,
2$\times r_{core}$, 2sqrt(2)$\times r_{core}$, 4$\times r_{core}$,
5$\times r_{core}$, 6$\times r_{core}$, 7$\times r_{core}$ where
$r_{core}= 1''$ and is the radius of aperture 3. Corrections to total
magnitudes for the aperture magnitudes of point sources  are also
supplied. The Kron, Petrosian and aperture magnitudes for both point
sources and extended objects are deblended of overlapping sources. Areal
profiles are given as the number of pixels above 8 thresholds.
Star-galaxy separation statistics, peak heights, sky levels and sky
variance are also calculated. A number of other parameters are given and
a full list is available at
\url{http://casu.ast.cam.ac.uk/surveys-projects/vst/technical/catalogue-
generation}

\subsection{Survey statistics}

\begin{figure}
\centering
%\resizebox{\hsize}{!}{\includegraphics{vst_Atlas_seeing_AB_only.png}}
%\resizebox{\hsize}{!}{\includegraphics{seeing.png}}
\resizebox{\hsize}{!}{\includegraphics{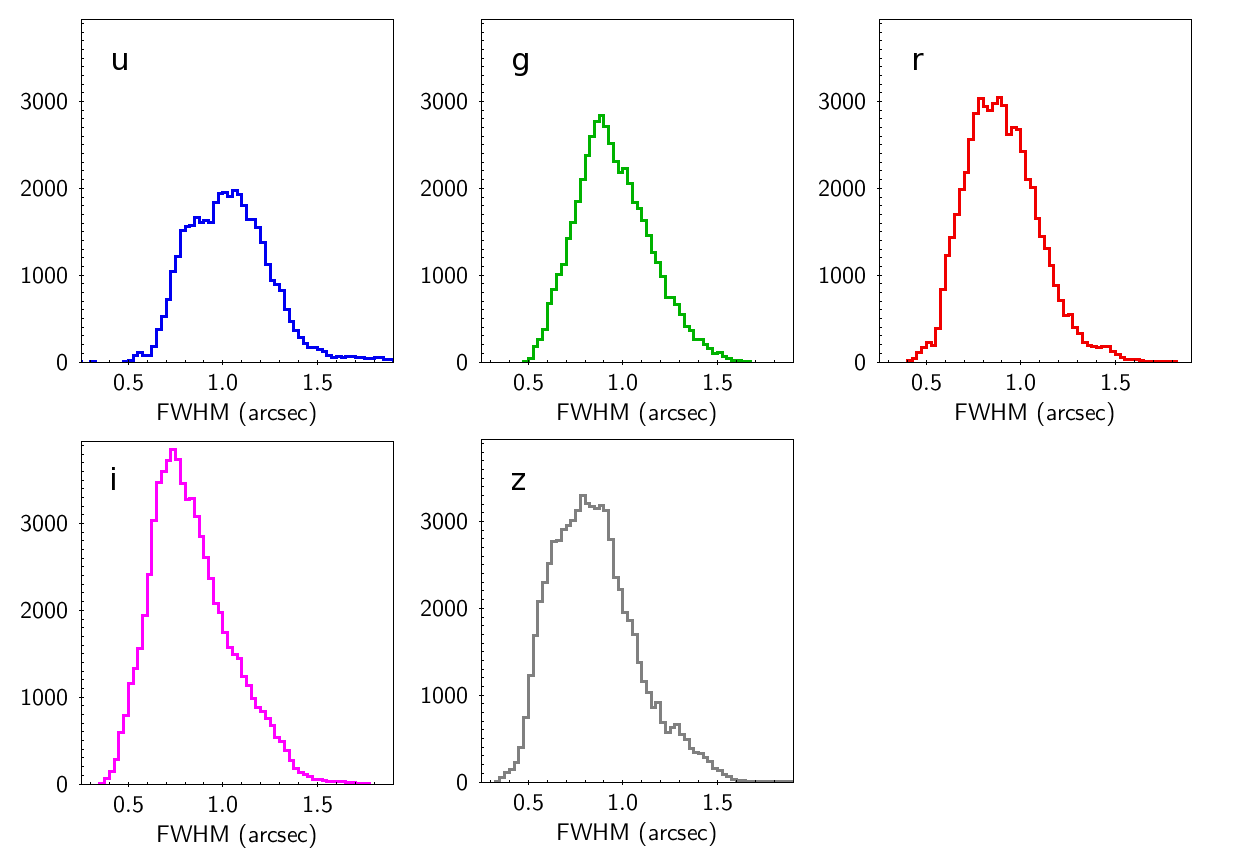}}
\caption{ATLAS FWHM seeing distributions in arcseconds. These are for ESO A and B grade tiles.}
\label{fig:seeing}
\end{figure}

\begin{figure}
\centering
%\resizebox{\hsize}{!}{\includegraphics{vst_Atlas_maglim_AB_only_inc_median.png}}
%\resizebox{\hsize}{!}{\includegraphics{5sig_maglim_mike_jrf_offsets.pdf}}
\resizebox{\hsize}{!}{\includegraphics{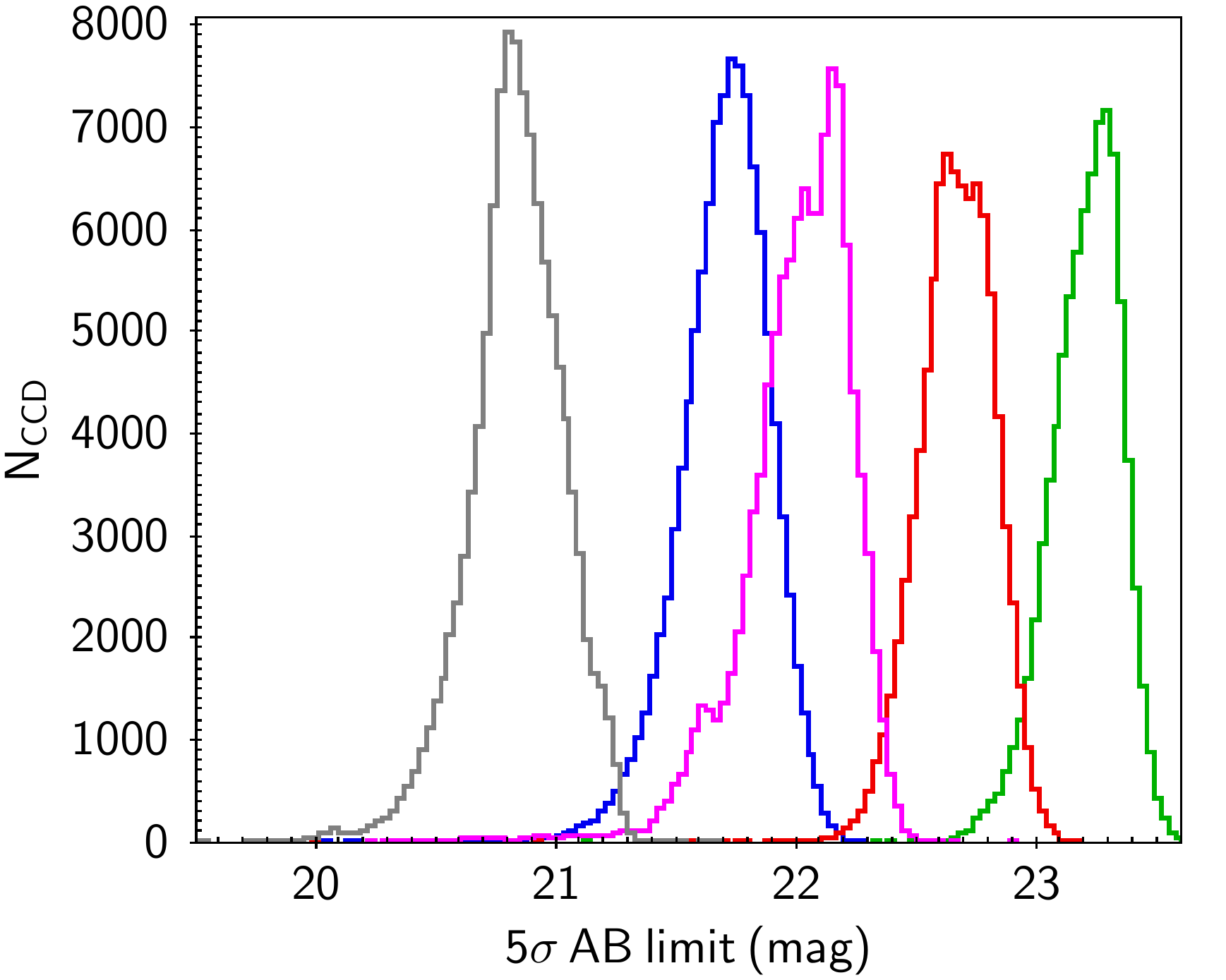}}
\caption{ATLAS 5$\sigma$ AB magnitude limit distributions for point
sources in  $u$ (blue), $g$ (green), $r$ (red), $i$ (purple) and  $z$
(grey). The median magnitude limits are given in Table \ref{tb:exp_time}
where they are compared to the equivalent SDSS limits.}
\label{fig:maglim}
\end{figure}

\begin{figure}
\centering
%\resizebox{\hsize}{!}{\includegraphics{sky_brightness_ab_17_23.png}}
%\resizebox{\hsize}{!}{\includegraphics{skybright_mike_jrf_offsets.pdf}}
\resizebox{\hsize}{!}{\includegraphics{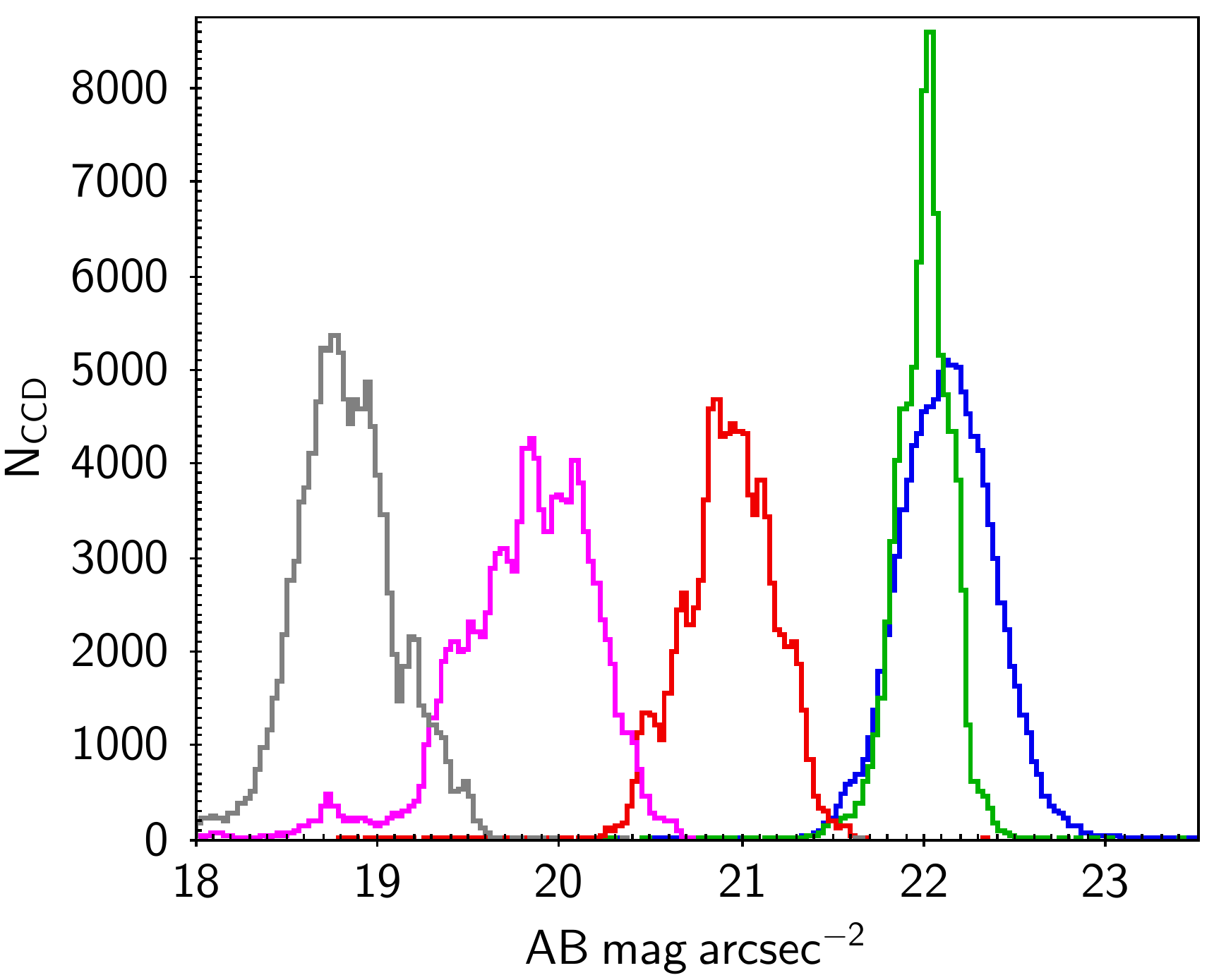}}
\caption{ATLAS sky brightness distributions in  $u$ (blue), $g$
(green), $r$ (red), $i$ (purple) and  $z$ (grey). The median sky
brightnesses are given in Table \ref{tb:exp_time} where they are
compared to their  SDSS equivalents.
}
\label{fig:skybright}
\end{figure}

The seeing distributions by passband are shown in Fig. \ref{fig:seeing}.
These show that the median seeing from the CASU measurements is
subarcsecond in the $riz$ bands,  and even at $ug$ only rises to
$\approx1''$ FWHM. The individual median seeings in each band are given
in Table \ref{tb:exp_time}.  These are significant improvements over
median SDSS seeing values e.g.  $1.''24$ FWHM in $r$ \citep{bramich12}.
The distribution of limiting magnitudes at the 5$\sigma$ detection level
by passband is shown in Fig. \ref{fig:maglim}. The median 5$\sigma$
stellar magnitude limit in a $1''$ radius aperture reaches
$r\approx22.7$. The median limits in all bands are also given in Table
\ref{tb:exp_time}. Also given there are ATLAS median magnitude limits
calculated in apertures of radius the SDSS FWHM seeing and these might
be thought more comparable to faint galaxy S/N limits. In Fig.
\ref{fig:skybright} we similarly show the distribution of sky
brightnesses and the medians are again given in Table \ref{tb:exp_time}.
Finally, in Table \ref{tb:exp_time}, we also present the count rate in
each ATLAS passband for a $20^m$ (AB) point source based on the
magnitude zeropoints.

%These are based on 5$\sigma$ detection limits for point sources assuming $1''$
%seeing and airmass 1.4 from \citet{goto02} who quote 22.3, 23.3, 23.1,
%22.3 and 20.8 mags for $ugriz$ in AB. We have then approximately
%converted these to the SDSS median seeing in Table \ref{tb:exp_time}
%assuming read/sky noise limited and to airmass 1.0.

%In a 60s exposure  a $u=21$ star records $\approx290$ electrons in the $u$ filter.
%This is $\approx2\times$ less than expected. In the other bands....

In Table \ref{tb:exp_time} we also present SDSS statistics for
comparison, including seeing, sky brightness and count rates. SDSS
seeing statistics are taken from \citet{bramich12} and sky brightness statistics are sourced from
\url{http://www.sdss2.org/dr7/}.  As already noted, in all bands median
ATLAS seeing is significantly better than that of SDSS.  Comparing SDSS
and ATLAS median sky brightnesses we note that the ATLAS sky
brightnesses are fainter than SDSS in $u$, $g$, $r$ and slightly
brighter in $i$ and $z$ due to ATLAS using grayer conditions for $i$,
$z$. The SDSS count rates are also given from \cite{stoughton02} eqn.
(1) assuming airmass 1.3,  as are the quoted ATLAS count rates. We see
that the ATLAS and SDSS count rates are similar with ATLAS having a
significant advantage only at $z$, the same conclusion as from the
throughput comparison in Fig. \ref{fig:filters}. The SDSS magnitude
limits have been calculated assuming these SDSS count rates and sky
brightnesses. They are again $5\sigma$ limits for point sources but now
measured in the radius of the quoted SDSS seeing (ie $\theta<1.''46$ in
$u$). This means that the ATLAS limits are $\approx0.^m25$ fainter in
$ugri$ due to better seeing and $0.^m7$ fainter in $z$, with this extra
advantage due to ATLAS' higher throughput in the $z$ band. The
comparison of these limits is more favourable to SDSS when the bigger
apertures are used to compute ATLAS 5$\sigma$ limits (Mag Lim (gal)).
These are more like the apertures appropriate for faint ATLAS galaxies.
Here ATLAS is within $0.15$mag of the SDSS limit in $ugrz$ but ATLAS has
a $0.^m31$ disadvantage in $i$ partly due to a brighter sky brightness.
Note that ATLAS is sky noise limited in $griz$ and read-out-noise
limited in $u$. We conclude that SDSS and ATLAS appear to have
comparable magnitude limits for galaxy photometry but that ATLAS has 
significantly fainter limits than SDSS for stellar photometry mainly
due to its better seeing.

%============================================================%
\section{Photometric scales and colour equations}
%=============================================================%

\subsection{ATLAS-Vega colour equations}

As noted above, the original photometric calibration (as used in ATLAS
DR1) was based on the limited number of standard fields observed on most
nights. This `ESO' calibration is in a VST Vega-like system and remains
available as an alternative calibration, despite the known issue with
occasional non-photometric conditions. Nevertheless, we list below the
linear colour equations used to convert the standards to an internal VST
system for the ESO Vega calibration :

$$u_{vst}  =  U + 0.035(U-B)$$
$$g_{vst}  =  B  -  0.405(B-V)$$
$$r_{vst}   =  R + 0.255(V-R)$$
$$i_{vst}    =  I  + 0.115(R-I)$$
$$z_{vst}   =  I   -  0.390(R-I)$$

\noindent As noted in Section \ref{sect:calibn}, we have since re-calculated  all
the zero-points based now on the illumination-correction fixed catalogues
while  computing  APASS AB  nightly and field-by-field magnitude  zero-points.

\subsection{ATLAS-SDSS colour equations}
\label{sect:SDSS_col_equ}
In the equatorial regions  there is a $\approx300$deg$^2$ overlap with
SDSS, split  between  the SGC ($\approx180$deg$^2$) and the NGC
($\approx120$deg$^2$). In the SGC the overlap is at
RA$\approx$22h40-03h, $-11<Dec<-9$ and in the NGC it is at
RA$\approx$10h00-15h30,$-3.5<Dec<-2$deg. 
We have  compared  $ugriz$ (AB) colours of SDSS and  ATLAS stars in the
120deg$^2$  NGC overlap region to define the colour equations between the two systems.  We use
PSF magnitudes for SDSS and aperture 3 (corrected to total) magnitudes for
ATLAS. The magnitude ranges used are $14<u<20$, $15<g<20$,
$15<r<20$, $14<i<19$ and $13.5<z<18.5$.

\begin{figure}
\centering
%\resizebox{\hsize}{!}{\includegraphics{sdss_colour_eqn.jpg}}
%\resizebox{\hsize}{!}{\includegraphics{colour_terms.jpg}}
\resizebox{\hsize}{!}{\includegraphics{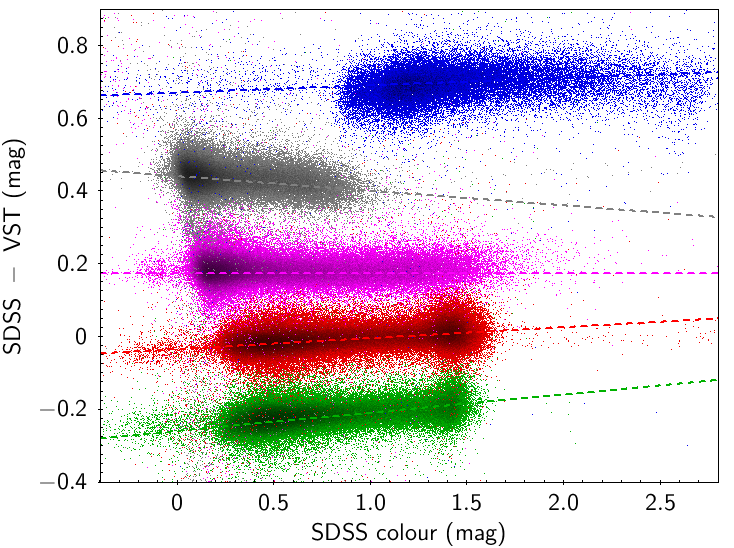}}
\caption{Colour equations between ATLAS aperture 3 AB magnitudes and
SDSS PSF magnitudes for stars in   $u$ (blue), $g$ (green), $r$ (red),
$i$ (purple) and  $z$ (grey). Here the ATLAS AB magnitudes are derived
from the APASS nightly zeropoints. The points have been displaced in the
ordinate  for clarity,  $g_{SDSS}-g$ by -0.2, $i_{SDSS}-i$ by +0.2,
$z_{SDSS}-z$ by +0.4 and $u_{SDSS}-u$ by +0.4. Apart from this, the
dashed lines are the equations given in Sect:\ref{sect:SDSS_col_equ}.
}
\label{fig:coleq}
\end{figure}

In Fig. \ref{fig:coleq} we show $u_{SDSS}-u:u-g$,
$g_{SDSS}-g:g-r$, $r_{SDSS}-r:g-r$, $i_{SDSS}-i:r-i$, $z_{SDSS}-z:i-z$.
We see that the colour terms in $u$ and $i$ are almost negligible but
there are small colour terms in $grz$. We find that:

$$u_{SDSS}=u_{VST} + 0.01(u-g)_{SDSS}  + 0.27$$
$$g_{SDSS}=g_{VST} + 0.05(g-r)_{SDSS}  - 0.06$$
$$r_{SDSS}=r_{VST} + 0.03(g-r)_{SDSS}  - 0.035$$
$$i_{SDSS}=i_{VST} - 0.025$$
$$z_{SDSS}=z_{VST} - 0.04(i-z)_{SDSS}  + 0.04$$

We note that the zeropoint offsets with SDSS are  small and due to
slight  differences in effective bandpass wavelengths. The exception  is
$u$ where the APASS AB zeropoints had to be fabricated from APASS $gri$.
This is  due to the non-linear (even double-valued for some star types)
transformation really required when going from $g,r$ to $u$.   We have
investigated if these offsets are caused by the method of magnitude
measurement and find that these results are robust. Note that this issue
in $u$ does not affect the AB  magnitudes in Table \ref{tb:exp_time} and
Figs. \ref{fig:apass_hist}, \ref{fig:maglim} and \ref{fig:skybright}
since these are based on the original ESO Vega calibration plus the
computed AB-Vega offsets given in Section \ref{sect:calibn}.

We find  good agreement in the colour coefficients with
the SDSS-VST $gri$ band conversions independently reported by the KiDS team at
\url{http://kids.strw.leidenuniv.nl/DR1/images/cterms.png}
as 

$$g_{SDSS}=g_{VST} + 0.0522(g-r)_{SDSS} $$
$$r_{SDSS}=r_{VST} + 0.0321(g-r)_{SDSS}$$
$$i_{SDSS}=i_{VST} + 0.0155(r-i)_{SDSS}$$
\noindent Typical extinction coefficients (mags)  for ATLAS  are 0.46($u$),
0.19($g$), 0.09($r$), 0.05($i$), 0.05($z$).
%\begin{figure}
%\centering
%\resizebox{\hsize}{!}{\includegraphics{standards_colour_eqn_i.jpg}}
%\caption{gri ESO standards observed in 4 ATLAS fields compared to ATLAS photometry.} 
%\label{fig:qso}
%\end{figure}

\subsection{ATLAS-SDSS comparison}

\begin{figure}
\centering
%\resizebox{\hsize}{!}{\includegraphics{sdsstest2.pdf}}
%\resizebox{\hsize}{!}{\includegraphics{r/sdss_vst_aper.png}}
%\resizebox{\hsize}{!}{\includegraphics{r/sdss_vst_petr.png}}
%\resizebox{\hsize}{!}{\includegraphics{r/sdss_vst_kron.png}}
\resizebox{\hsize}{!}{\includegraphics{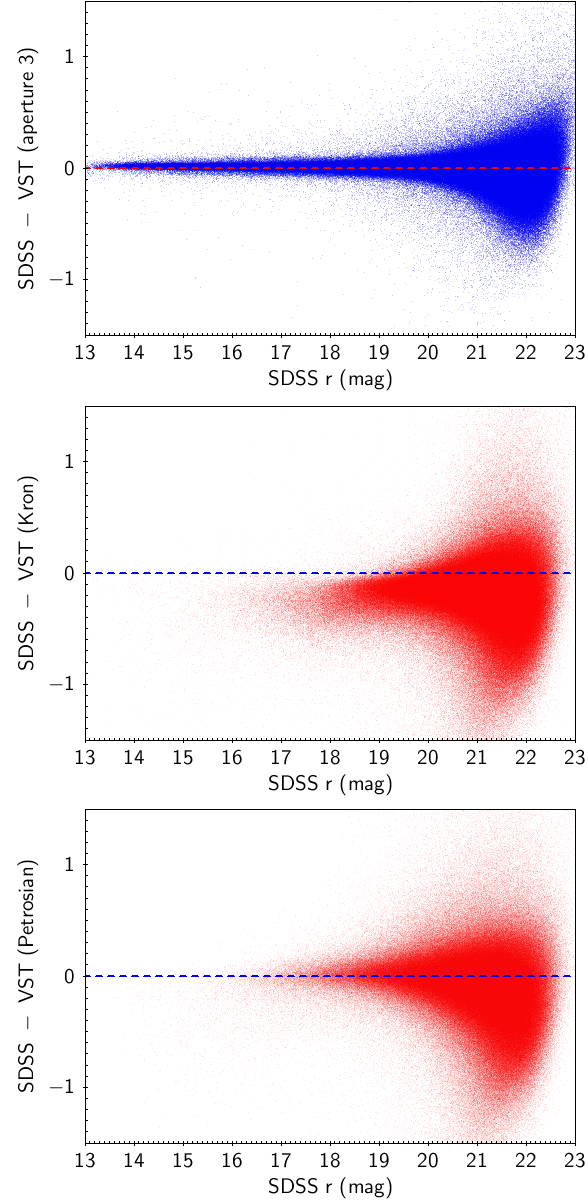}}
\caption{SDSS NGC overlap comparisons of ATLAS (APASS nightly)  aperture (corrected to total), 
Kron and  Petrosian AB magnitudes (corrected for colour terms as in Section
\ref{sect:SDSS_col_equ}) for stars (blue) and  galaxies (red) with  SDSS
model magnitudes in both cases. No corrections to total magnitudes  have
been applied to the Kron and  Petrosian magnitudes.}
\label{fig:sdsscomp}
\end{figure}

We next  compare linearity of the ATLAS photometry (calibrated via
nightly APASS and after the above colour corrections), with SDSS in the
$\approx120$deg$^2$  NGC overlap area. We shall see later that there are
still some area-dependent zeropoint problems with the nightly APASS
calibrations but here we will focus on the $r$-band where the
field-field problems are less. In the top panel of Fig.
\ref{fig:sdsscomp} we plot ATLAS aperture magnitudes versus model
magnitudes for SDSS stars. The relation is consistent with linearity
over the full $13<r<22.5$ mag range.  In the middle panel of Fig.
\ref{fig:sdsscomp} we plot ATLAS Kron magnitudes against SDSS model
magnitudes for galaxies. An offset to total of -0.15mag for galaxies is
suggested for ATLAS.   The relation   appears noisier than  the stars 
as expected but with no  evidence of a scale error to $r\approx22$. In
the bottom  panel of Fig. \ref{fig:sdsscomp} we similarly plot ATLAS
Petrosian magnitudes for galaxies against model magnitudes for SDSS. We
see a reasonably linear, if noisier, relation. In the $22<r<22.5$ bin
the rms error reaches $\pm0.2$, $\pm0.38$ and $\pm0.38$ mag for
aperture, Petrosian and Kron magnitudes and here we might expect roughly
equal contributions from ATLAS and SDSS. Overall, for VST ATLAS, we
therefore recommend that aperture magnitudes (aper$_3$) are used for
stars while Petrosian or Kron magnitudes are used for galaxies.

These results for the $r$ band are typical of the results for the other 4 bands.
These can be found at \url{http://astro.dur.ac.uk/Cosmology/vstatlas/tests/}.

%\begin{figure}
%\centering
%\resizebox{\hsize}{!}{\includegraphics{r_kron.jpg}}
%\caption{Comparisons of ATLAS Kron magnitudes for stars (blue) and  galaxies(red)
%with  WHDF Kron magnitude.}
%\label{fig:whdfcomp}
%\end{figure}

\subsection{ATLAS-Stripe 82 (WHDF field) comparison}

Because SDSS only reaches a similar depth to ATLAS, it is not clear
which survey is dominating the error spread at faint magnitudes. We have
therefore also  compared ATLAS photometry for galaxies and stars  with
the much deeper Stripe 82 (S82) photometry which lies within a 1deg$^2$,
specially observed, ATLAS area which also contains the William Herschel
Deep Field (Metcalfe et al 2005) and again its deeper multi-epoch
photometry allows a comparison to check the magnitude dependent error of
the ATLAS photometry.

We note that although the WHDF ATLAS data has the same exposure as the
ATLAS survey data and was taken in dark time, the seeing particularly in
$u$ and $g$ is significantly worse than average (see Table
\ref{tb:atlas_whdf}). Indeed, the $5\sigma$ point source magnitude
limits in all bands are brighter than the ATLAS average from $\approx0.2$mag in
$riz$, to 0.64mag in $g$ and 0.54mag in $u$ (see Table
\ref{tb:atlas_whdf}). Nevertheless we can still make approximate checks of these brighter
than usual  WHDF ATLAS mag limits  by comparison to deeper S82 data.

\begin{table}
\centering
\begin{tabular}{cccccc}
\hline\hline
Band & $u$ & $g$ & $r$ & $i$ & $z$ \\
\hline\hline
Exposure       & $2\times60$s & $2\times50$s & $2\times45$s & $2\times45$s & $2\times45$s \\
Airmass        & $1.52$   & $1.46$   & $1.16$   & $1.14$   & $1.13$ \\
Seeing         & $1.''27$ & $1.''36$ & $1.''04$ & $0.''99$ & $1.''06$ \\
ATLAS Seeing   & $1.''02$ & $0.''95$ & $0.''90$ & $0.''81$ & $0.''84$ \\
%Mag Lim       & $21.56$ & $22.50$ & $22.50$ & $21.89$ & $20.63$ \\ AB-Vega=1.0, -0.1, 0.2, 0.4, 0.5
Mag Lim        & $21.45$ & $22.50$ & $22.46$ & $21.85$ & $20.65$ \\
%ATLAS Mag Lim & $22.10$ & $23.14$ & $22.71$ & $22.03$ & $20.85$ \\  AB-Vega=1.0, -0.1, 0.2, 0.4, 0.5
ATLAS Mag Lim  & $21.99$ & $23.14$ & $22.67$ & $21.99$ & $20.87$ \\
Sky Brightness & $21.56$ & $21.64$ & $20.77$ & $19.96$ & $18.80$ \\
ATLAS Sky Bri  & $22.34$ & $21.90$ & $20.92$ & $19.78$ & $18.85$ \\
\hline
%\hline\hline
\end{tabular}
\caption[]{ATLAS WHDF field seeing, $5\sigma$ point source magnitude limits and sky brightnesses, compared to ATLAS 
median parameters from Table \ref{tb:exp_time}. All magnitudes are quoted in the AB system.}
\label{tb:atlas_whdf}
\end{table}

\begin{figure}
\centering
%\resizebox{\hsize}{!}{\includegraphics{r_aper_stripe82_new_v2.png}}
%\resizebox{\hsize}{!}{\includegraphics{r_petr_stripe82_new.png}}
%\resizebox{\hsize}{!}{\includegraphics{r_kron_stripe82_new.png}}
\resizebox{\hsize}{!}{\includegraphics{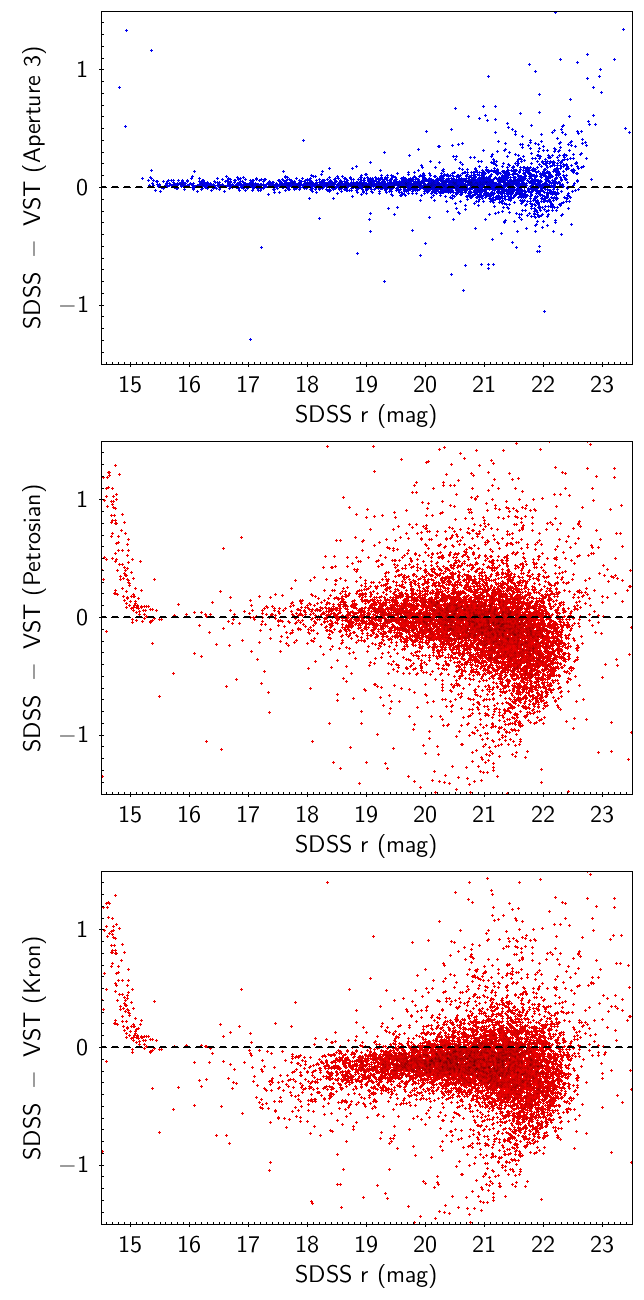}}
\caption{(top) SDSS Stripe 82 model magnitude-ATLAS aperture magnitude
r-band comparison in the WHDF field for stars (blue). (middle) Same
comparison now based on ATLAS Petrosian magnitudes for galaxies (red).
(bottom) Same comparison now based on ATLAS Kron magnitudes for  
galaxies (red). In the two lower panels, saturation in S82 magnitudes 
of mis-classified stars causes the non-linearity at $r<15.5$.
}
\label{fig:s82_atlas}
\end{figure}

So in Fig. \ref{fig:s82_atlas} we now show in the top panel the
comparison between S82 model  magnitudes and ATLAS aperture 3 magnitudes
for point sources. Again the  relation is linear with a low dispersion
of $\pm0.15$mag at the $5\sigma$ WHDF ATLAS  limit of $r=22.46$ from
Table \ref{tb:atlas_whdf}. Indeed, ATLAS aperture magnitudes in all
bands show good agreement with the S82 magnitudes to their respective
5$\sigma$ limits from Table \ref{tb:atlas_whdf}, confirming that the
accuracy of these limits for point sources (see 
\url{http://astro.dur.ac.uk/Cosmology/vstatlas/tests}). The middle and bottom 
panels show the comparison between S82 model magnitudes  and ATLAS
Petrosian/Kron magnitudes now for SDSS galaxies. Both  Petrosian and
Kron magnitudes are reasonably linear but Petrosian appears to have a
higher dispersion at $r\approx21.5$ of $\pm0.2$mag as opposed to
$\pm0.15$mag for Kron. On this basis the preferred ATLAS magnitude for
galaxy photometry again appears to be Kron magnitudes. 

%=
%=============================================================%
\section{Zero-point comparisons with APASS and SDSS}
%=============================================================%

\subsection{APASS v ATLAS}  
\label{sect:apass_atlas}

\begin{figure}
\centering
%\resizebox{\hsize}{!}{\includegraphics{apass_new_u.jpg}}
%\resizebox{\hsize}{!}{\includegraphics{apass_new_g.jpg}}
%\resizebox{\hsize}{!}{\includegraphics{apass_new_r.jpg}}
%\resizebox{\hsize}{!}{\includegraphics{apass_new_i.jpg}}
\resizebox{\hsize}{!}{\includegraphics{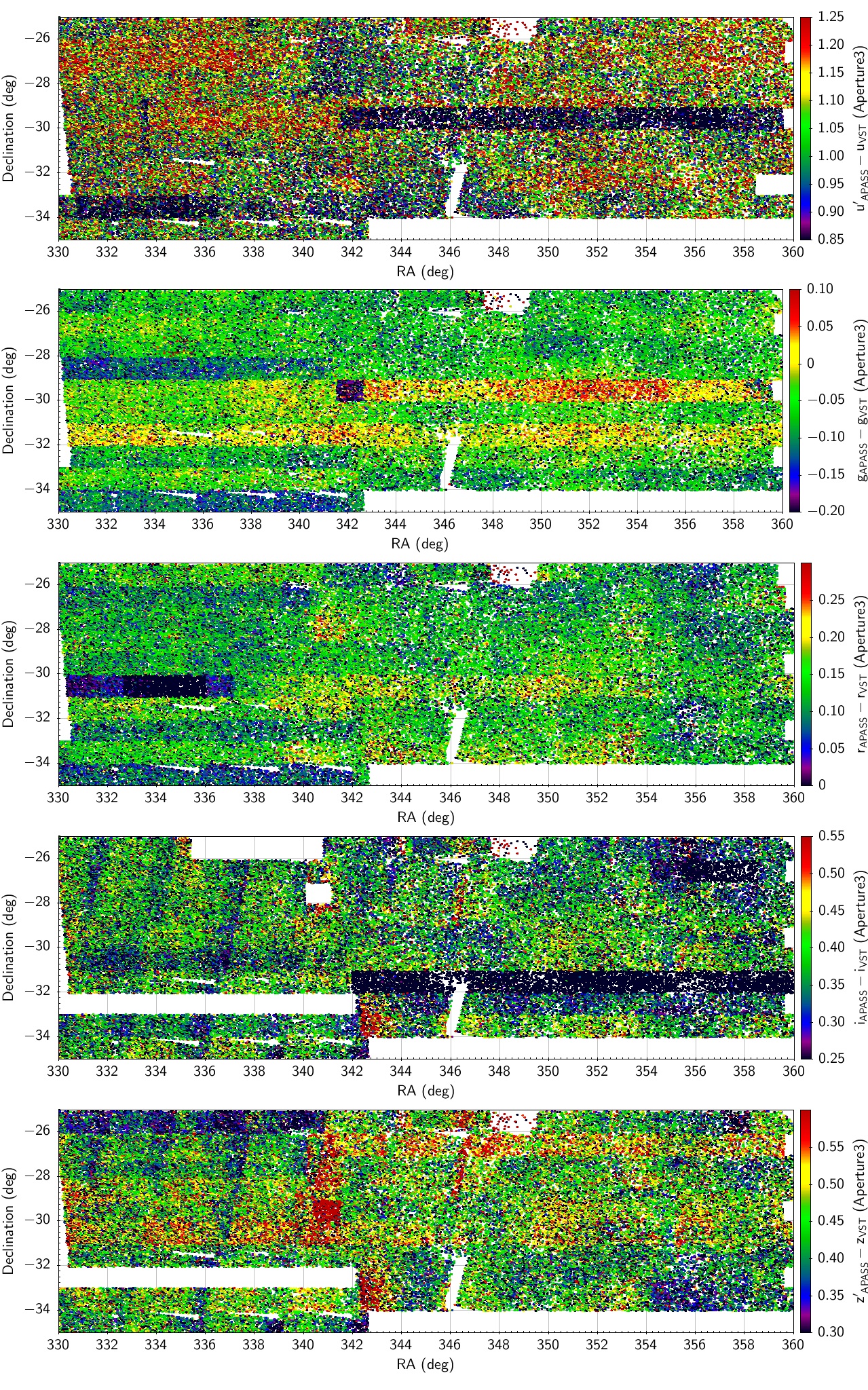}}
\caption{From top to bottom, APASS-ESO ATLAS comparisons in $u$ (blue), $g$ (green), $r$
(red), $i$ (purple), $z$ (grey). APASS mags  are in AB and
ATLAS in Vega. These are for ATLAS stacks of ESO A,B grade tiles.}
\label{fig:apass_area}
\end{figure}

We now move away from checks of the linearity of ATLAS photometry to
address in more detail the question of the consistency of the ATLAS
zeropoint calibration as discussed in Section \ref{sect:calibn} over
wider areas of the survey. Maps of ESO ATLAS-APASS residuals in a
$\approx300$ deg$^2$  area centred on RA$\approx23$h, Dec$\approx-30$deg
are shown in Fig. \ref{fig:apass_area}. The ATLAS data is calibrated
from the ESO standards as used for the DR1 data release. The offsets in
$u$ are obtained by extrapolating to $u$ from APASS $B,V$ and the $z$
offsets are from interpolating APASS $r$ and $i$ measurements. Offsets
are  seen that are clearly due to ATLAS/ESO calibration issues
(generally in blocks of $\sim17\times1$deg, corresponding to a single
concatenation of tiles).  Clearly some within field residuals are seen,
particularly in $i$ (and $z$), that do not show any ATLAS
characteristics and we believe that  these may be due to problems in the
APASS $i$-band (and its extrapolation to $z$). The distribution of
APASS-ATLAS (tile-by-tile) residuals for $ugriz$ over a larger area is
shown in Fig. \ref{fig:apass_hist} and the standard deviations are
$\pm0.^m060$ ($u$), $\pm0.^m060$ ($g$), $\pm0.^m045$ ($r$), $\pm
0.^m038$ ($i$), $\pm 0.^m080$ ($z$).

\subsection{ATLAS-SDSS Equatorial comparison}
\label{sect:atlas_sdss}
\begin{figure}
\centering
\resizebox{\hsize}{!}{\includegraphics{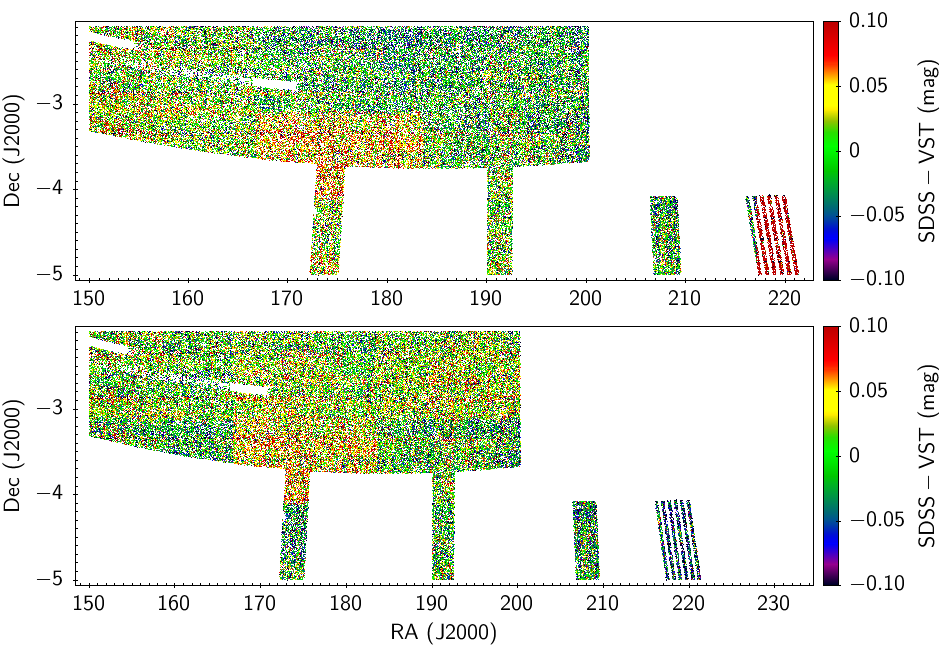}}
\caption{SDSS-ATLAS comparison in $u$ with ATLAS/ESO-SDSS (top)
and ATLAS/APASS nightly-SDSS (bottom).  Aperture 
3 mags are used for ATLAS and PSF magnitudes for SDSS.}
\label{fig:sdss_area_u}
\end{figure}

\begin{figure}
\centering
%\resizebox{\hsize}{!}{\includegraphics{g_all.png}}
\resizebox{\hsize}{!}{\includegraphics{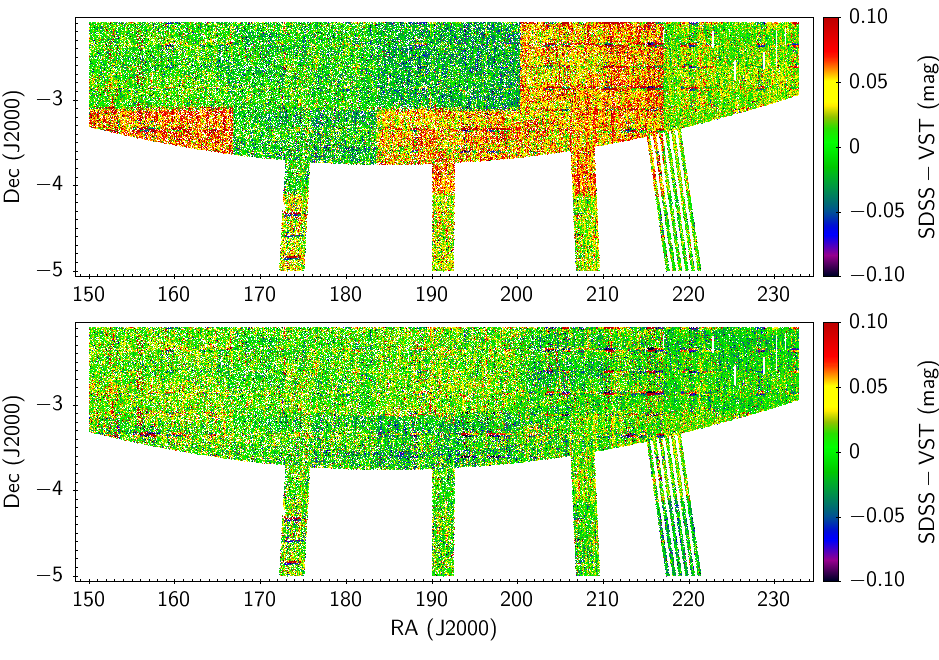}}
\caption{SDSS-ATLAS comparison in $g$ with ATLAS/ESO-SDSS (top)
 and ATLAS/APASS nightly-SDSS (bottom).  Aperture 
3 mags are used for ATLAS and PSF magnitudes for SDSS.}
\label{fig:sdss_area_g}
\end{figure}

\begin{figure}
\centering
\resizebox{\hsize}{!}{\includegraphics{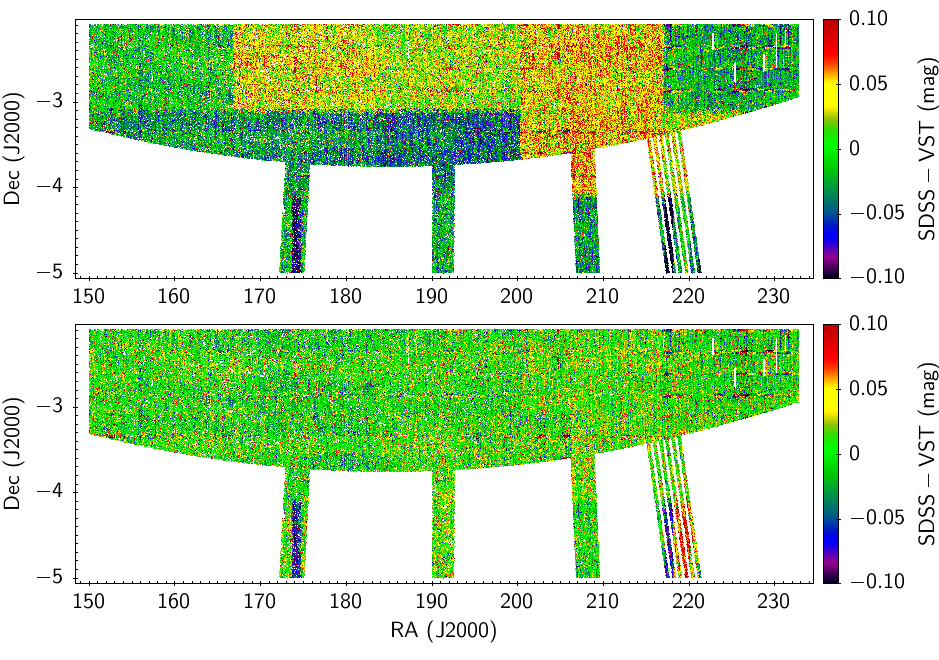}}
\caption{SDSS-ATLAS comparison in $r$ with ATLAS/ESO-SDSS (top)
and ATLAS/APASS nightly-SDSS (bottom).  Aperture 
3 mags are used for ATLAS and PSF magnitudes for SDSS.}
\label{fig:sdss_area_r}
\end{figure}

\begin{figure}
\centering
\resizebox{\hsize}{!}{\includegraphics{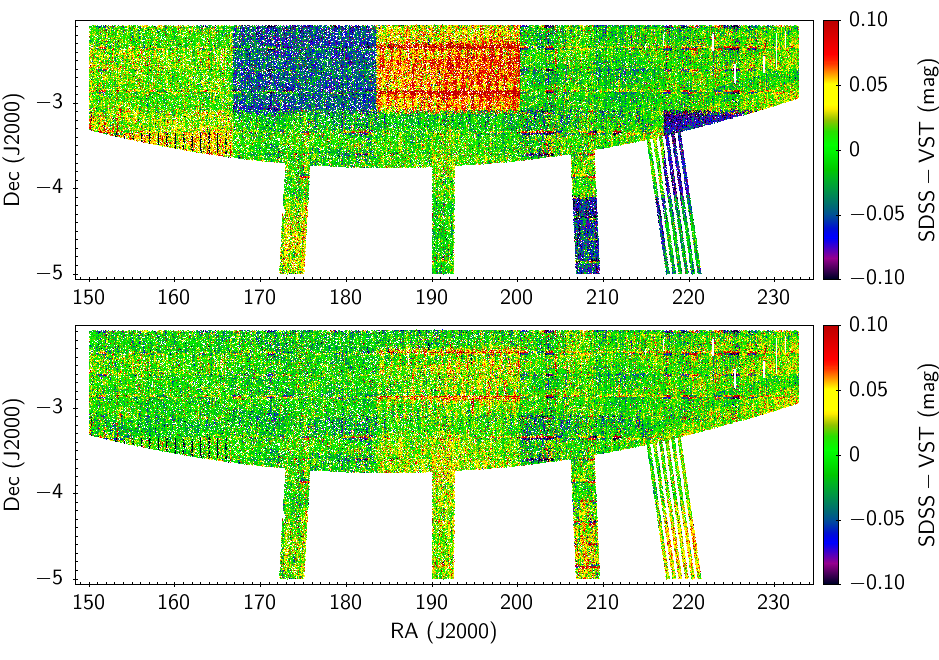}}
\caption{SDSS-ATLAS comparison in $i$ with ATLAS/ESO-SDSS
(top) and ATLAS/APASS nightly-SDSS (bottom).  Aperture 3 mags are used
for ATLAS and PSF magnitudes for SDSS.}
\label{fig:sdss_area_i}
\end{figure}

\begin{figure}
\centering
\resizebox{\hsize}{!}{\includegraphics{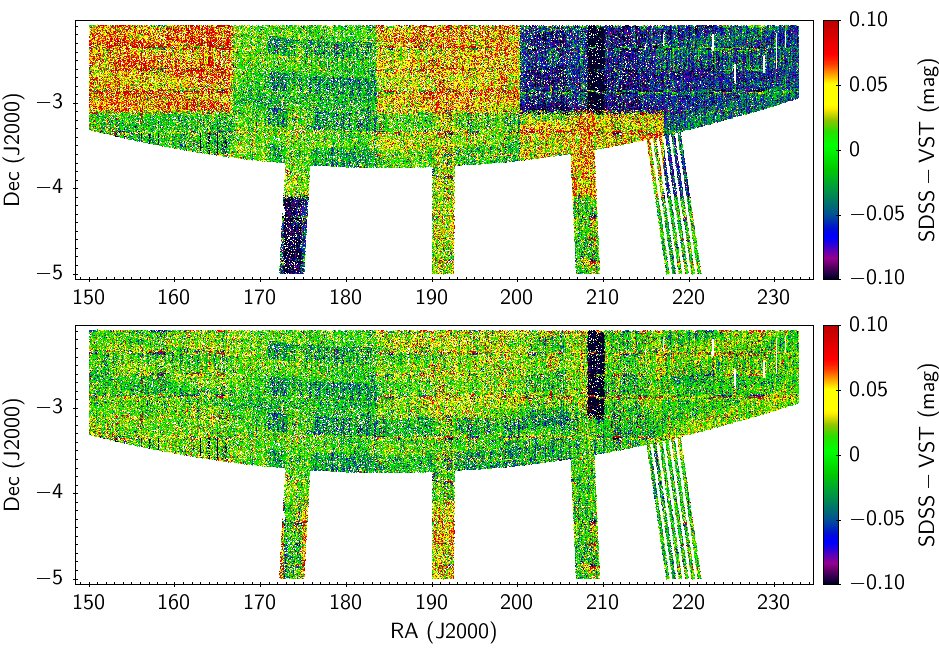}}
\caption{SDSS-ATLAS comparison in $z$ with ATLAS/ESO-SDSS (top)
 and ATLAS/APASS nightly-SDSS (bottom).  Aperture 
3 mags are used for ATLAS and PSF magnitudes for SDSS.}
\label{fig:sdss_area_z}
\end{figure}

Figs. \ref{fig:sdss_area_u}-\ref{fig:sdss_area_z} show the residuals
between the SDSS data for stars in the NGC overlap area at
10h$<$RA$<$15h30 and $-3.5<$Dec$<-2$deg assuming,  in turn, the ESO
calibration  and the APASS nightly calibration for ATLAS. The SDSS
magnitude ranges used are $14<u<20$, $15<g<20$, $15<r<20$, $14<i<19$ and
$13.5<z<18.5$.  The standard SDSS calibration was used here for
simplicity, rather than the uber-calibration of \citet{pad08}. As with
the APASS comparison in Section \ref{sect:apass_atlas}, the ESO
calibration clearly shows $\pm0.^m05$ offsets on the scale of single
concatenations. These are effectively removed by using the APASS nightly
calibration. The ESO calibration also has larger residuals  with clear
VST nightly correlations. Some nights have no useable ESO standards and
on these a long term average default value is used, perhaps  explaining 
some of the stripes. However, even the APASS nightly calibration still
has occasional issues with individual fields as can be seen in the $z$
band image. Also, we already noted that the APASS $i$ (and $z$) band
photometry still has some issue from the ATLAS-APASS comparison above. 

\begin{figure}
\centering
%\resizebox{\hsize}{!}{\includegraphics{sdss_apass_atlas_offset.png}}
\resizebox{\hsize}{!}{\includegraphics{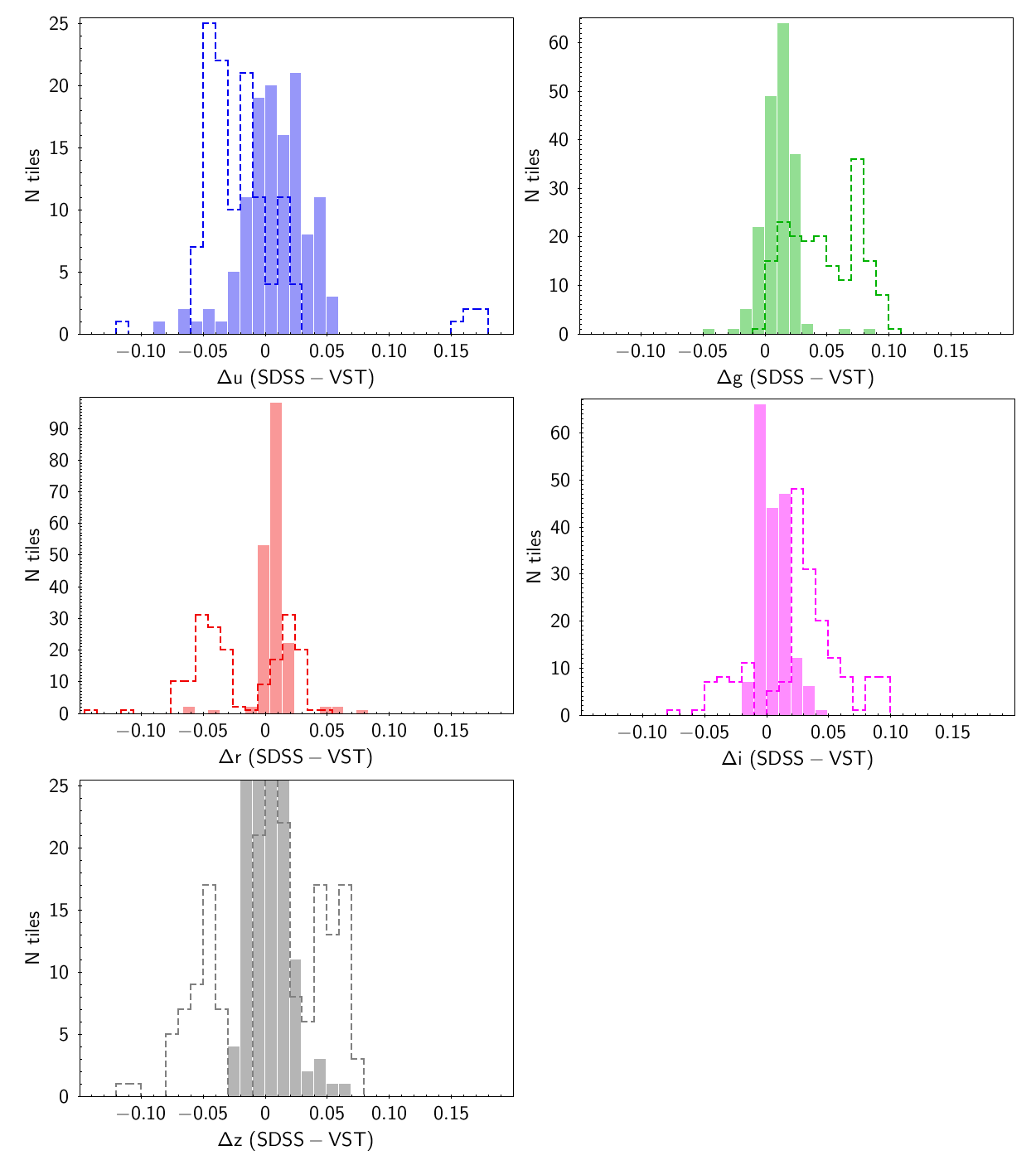}}
\caption{SDSS - ATLAS magnitude residuals for point sources in the SDSS
overlap area. The dashed histograms refer to ATLAS data with  ESO
zeropoints. The solid histograms refer to ATLAS data with APASS nightly
zeropoints. The magnitude ranges used are $14<u<20$, $15<g<20$,
$15<r<20$, $14<i<19$ and $13.5<z<18.5$.}
\label{fig:sdss_apass_offsets}
\end{figure}

The distributions of SDSS-ATLAS  residuals   for individual stars in
$ugriz$ for ESO and APASS calibrations are compared  in Fig.
\ref{fig:sdss_apass_offsets}. Generally the APASS nightly calibrated
data gives the smallest residuals with SDSS. The ESO calibration  and
the APASS individual fields calibration (not shown) appear to give
larger residuals, in the latter case presumably due to statistical
errors with the calibration based on fewer APASS stars. The ATLAS APASS
nightly-SDSS  mag standard deviations, measured on a tile-by-tile basis
to reduce the effect of individual star random errors, are $\pm0.035$,
$\pm0.013$, $\pm0.013$, $\pm0.012$ and $\pm0.055$ mag in $ugriz$, in
most bands a significant improvement over the ESO ATLAS - SDSS
comparison ($\pm0.045$, $\pm0.027$, $\pm0.037$, $\pm0.035$ and
$\pm0.073$ in $ugriz$). These standard deviations were measured on the
basis of 183 tiles in $griz$ and 112 tiles in $u$. In both cases, the
$z$ band result  is affected by the presence of a single tile whose tile
average is 0.66mag offset from the mean. Nevertheless in all bands the
improvement gained from the APASS nightly calibration is clear.

%The ATLAS APASS nightly-SDSS clipped ($\pm0.2$) mag standard deviations
%are $\pm0.054$, $\pm0.035$, $\pm0.035$, $\pm0.033$ and $\pm0.045$ mag in
%$ugriz$, in most bands a significant improvement over the ESO ATLAS -
%SDSS comparison ($\pm0.057$, $\pm0.043$, $\pm0.048$, $\pm0.046$ and
%$\pm0.061$ in $ugriz$). Clearly, these standard deviations, in the
%absence of any calibration errors, would be dominated by the average
%error on stellar photometry at the various magnitude limits.

Ultimately, for the final release, we shall calibrate  ATLAS globally by
incorporating the information from  the $2'$ overlaps in the RA and
Dec directions on each field (Findlay et al. 2014b in prep.)

%\begin{figure}
%\centering
%\resizebox{\hsize}{!}{\includegraphics{apass_offset_hist.jpg}}
%\caption{SDSS-APASS nightly offset histogram}
%\label{fig:sdss_offsets}
%\end{figure}

\subsection{Star-galaxy separation comparisons}
\label{sect:sg}
\begin{figure}
\centering
%\resizebox{\hsize}{!}{\includegraphics{sg_comparison.png}}
\resizebox{\hsize}{!}{\includegraphics{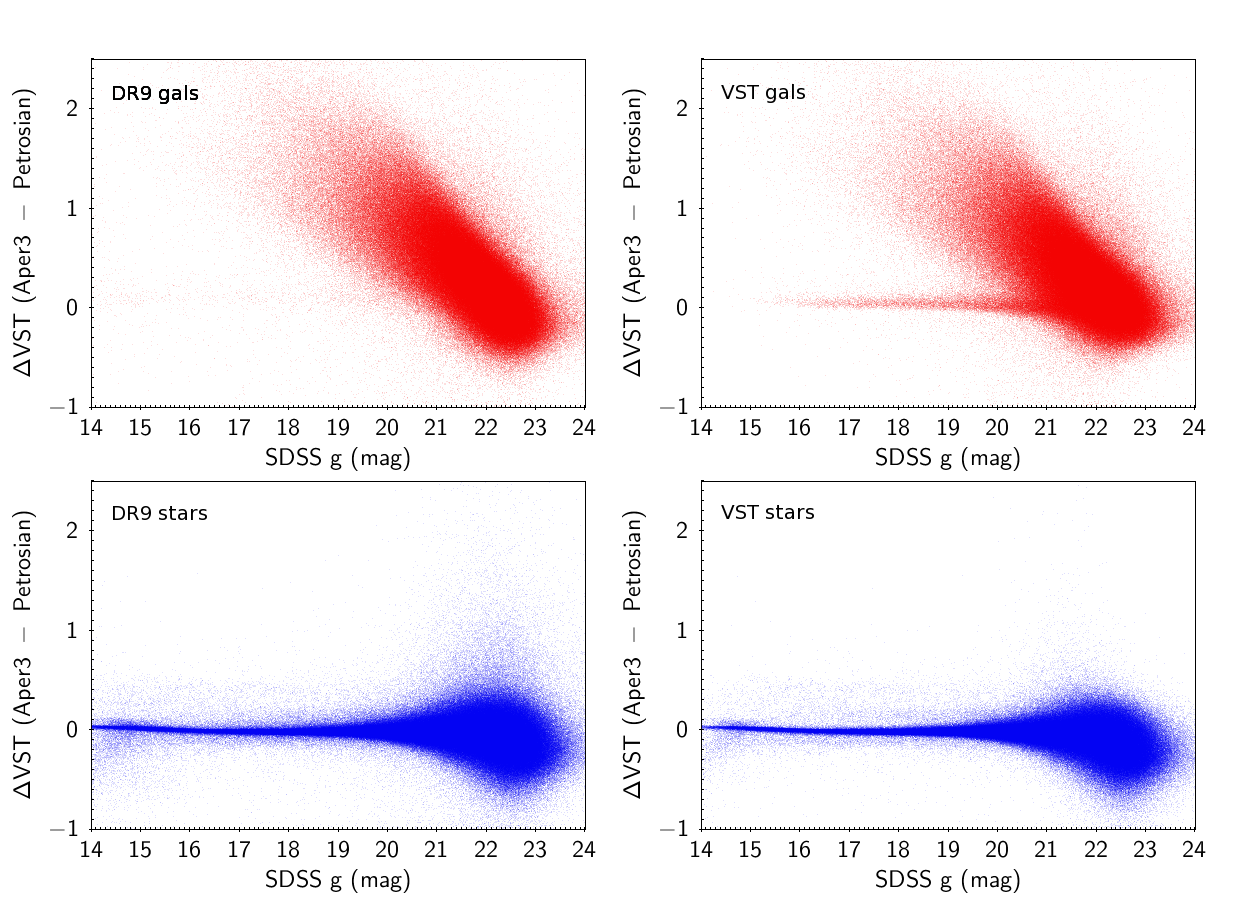}}
\caption{ Star-galaxy separation comparison. {\it Upper}: galaxies (red)
{\it Lower}: stars (blue). ATLAS stars mis-classified as galaxies
according to SDSS in the top right panel are due to problems in
partially covered interchip gaps.  }
\label{fig:sgcomp}
\end{figure}

\begin{figure}
\centering
\resizebox{\hsize}{!}{\includegraphics{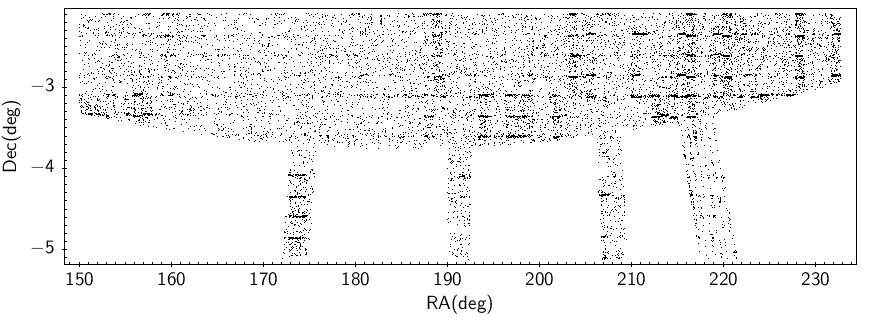}}
\resizebox{\hsize}{!}{\includegraphics{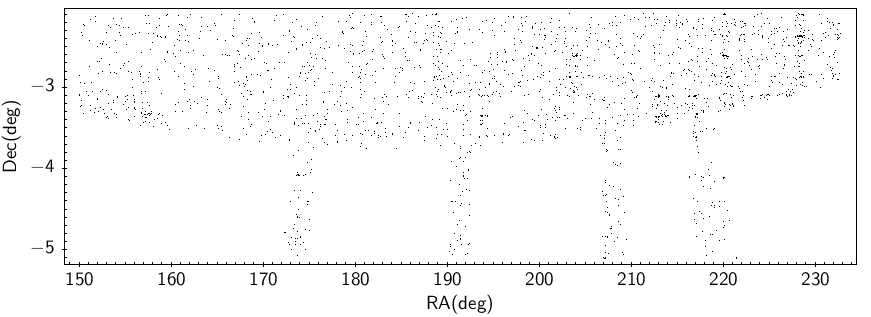}}
\caption{{\it Upper} Map showing ATLAS classified  galaxies which are stars according to
SDSS, down to $g\approx20.5$. These are found particularly in areas corresponding to interchip gaps. {\it Lower}:
Same as above except the additional  star-galaxy separation method
described in Section  \ref{sect:sg} has now been applied to the ATLAS
galaxies.}
\label{fig:misclass_g}
\end{figure}

We now use the same NGC SDSS DR9 overlap area to compare ATLAS
star-galaxy separation with that of SDSS. The results are shown in  Fig.
\ref{fig:sgcomp}. Comparing top left and top right panels in the ATLAS
Aper3-Petrosian : $g$ magnitude star-galaxy separation plane, we see
that the ATLAS star-galaxy separation statistic  has included some SDSS
stars just above the stellar loci shown in the lower panels. These
misclassified ATLAS  stars have come from areas frequently coinciding
with interchip gaps as can be seen in the upper panel of Fig.
\ref{fig:misclass_g}. The problem arises when the image quality varies
slightly between the 2 sub-exposures and so  the stellar locus is offset
in the interchip areas when there is only one sub-exposure contributing
to the image. Depending on the direction of the shift, this can result
in stars being classified as galaxies. This problem will clearly be
present in the ATLAS DR1 data. However, it is easy to fix this problem
by plotting $\Delta$ VST (Aper$_3$-Petrosian) versus VST Petrosian for
ATLAS classified galaxies and then remove the misclassified stars by a
magnitude independent cut. This has the effect  of cleaning up the
mis-classified stars as shown in the lower panel of Fig. 
\ref{fig:misclass_g}.

%=============================================================%
\section{Galaxy and Star number counts}
\begin{figure*}
\centering
\resizebox{\hsize}{!}{\includegraphics{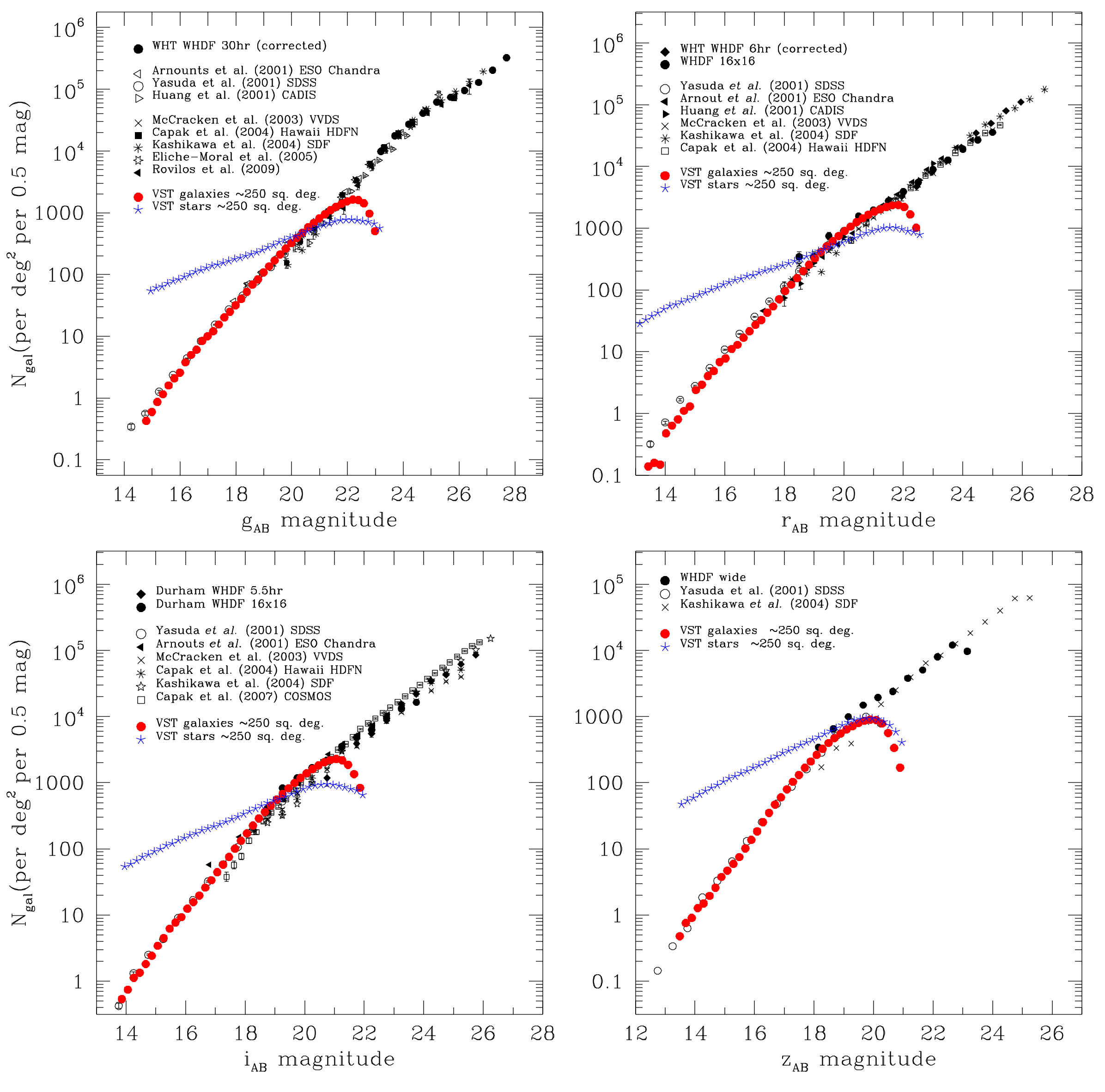}}
%\includegraphics{vst_nm.png}
%\resizebox{\hsize}{!}{\includegraphics{gc.jpg},\includegraphics{rc.jpg}}
%\resizebox{\hsize}{!}{\includegraphics{ic.jpg},\includegraphics{zc.jpg}}
%\resizebox{\hsize}{!}{\includegraphics{uc.jpg}}
\caption{ATLAS galaxy and star counts compared to SDSS and other galaxy counts. Both stars and galaxies
are measured in Kron magnitudes, brightened by $0.^m15$ to take them to total.}
\label{fig:counts}
\end{figure*}

%=============================================================%
Adopting the corrected star - galaxy classifier, we next calculate  the
galaxy (and star)  counts from 250deg$^2$ in the $-40<Dec<-25$ deg range
in the SGC which includes the SGP. Previously 2MASS counts have
suggested evidence  for a local underdensity in the SGP region. Here we
simply focus on the counts to use their turnover to indicate the rough
completeness limits of the survey. In Fig. \ref{fig:counts} we compare
these to a variety of previous galaxy counts compiled by
\citet{metcalfe01}. 2MASS counts suggest evidence  for a local
underdensity in the SGP region \citep{whitbourn14b} but a fuller
investigation of this issue using ATLAS data in a bigger area  will be
presented by  Mackenzie et al (2014, in prep.). We see that the galaxy
counts agree very well in most bands with other authors until the counts
turn over due to incompleteness at faint magnitudes. This happens at AB
mags $u\approx20$, $g\approx22.5$, $r\approx22$, $i\approx21$ and
$z\approx20$. These are in reasonable agreement with the Mag Lim (gal)
5$\sigma$ limits for ATLAS given in Table \ref{tb:exp_time}. Star counts
turn over at about the same limits but this is more dictated by the
star-galaxy separation algorithm and is less related to more fundamental
signal-to-noise considerations.

%=============================================================%
\section{Access to ATLAS data}
%=============================================================%

The ESO Science Archive provides a repository for data products released by
the ESO Public Survey teams. Images and catalogues from ATLAS Data Release 1
can be retrieved from
\url{http://archive.eso.org/wdb/wdb/adp/phase3_main/form?phase3_collection=ATLAS&r
elease_tag=1}.

ATLAS data are also available through the OmegaCAM Science Archive (OSA,
\url{http://osa.roe.ac.uk}), curated by the Wide-Field Astronomy Unit (WFAU) at
the University of Edinburgh. Similar in design to the science archives
developed by WFAU for sky survey data from UKIRT/WFCAM \citep{hambly08}
and VISTA \citep{cross12}, the OSA provides a range of access methods to
both image and catalogue data products. It stores ATLAS catalogue data and
image metadata in a relational database management system, which can be
queried through the OSA web interface or using the Virtual Observatory (VO)
Table Access Protocol \citep{dowler11}. Query results can be
viewed within the OSA web interface or sent directly from it to VO-compliant
desktop tools like Aladin \citep{bonnarel00} and TOPCAT \citep{taylor05},
using the Simple Application Message Protocol \citep{taylor12} standard.
The OSA also integrates ATLAS catalogue data with a number of sky survey
datasets, including SDSS, 2MASS, VHS and WISE, and provides the ATLAS
consortium with access to proprietary data, as well as supporting use of
ATLAS public data releases by the wider astronomical community.

%=============================================================%
\section{Conclusions}
%=============================================================%
We have described the basic characteristics of VST ATLAS. We find it
meets its basic specification of comprising a `Southern Sloan'. This
applies both  to  the median  $5\sigma$ magnitude limits for
point sources which are at least as deep as SDSS and to the median
sky brightnesses which are comparable to SDSS. But in terms of its
spatial  resolution, the ATLAS survey presents essentially sub-arcsecond
$0.''8-1.''0$ median seeing compared to the $1.''2-1.''5$ median imaging of SDSS.

We have derived the colour equations relating ATLAS stellar photometry
to SDSS photometry. We find there are small but non-negligible colour
terms, particularly in $grz$.

We have demonstrated the linearity of the ATLAS photometry by comparison
in the 120deg$^2$ NGC overlap area with SDSS. We recommend that $1''$ radius
(aperture 3) apertures are used for stellar photometry. For galaxies,
ATLAS Kron magnitudes seem to give the most linear relations with SDSS
model magnitudes. The same conclusions are found by comparing ATLAS and
Stripe 82 photometry in the WHDF field. Here we found that the rms
dispersions reached $\pm0.05$ mag for stars at $r<22$ and $\pm0.15$mag 
for Kron galaxy magnitudes at $r<21.5$mag

The current zeropoints assumed for $ugriz$ come from APASS stellar
photometry applied nightly and appear good to $\approx\pm0.02$mag by comparison
with SDSS data in overlap areas. We have investigated how area dependent
these zeropoints are in the SDSS overlap area and find there are still
small areas with residual errors. In the final global calibration we
shall use the $2'$ overlaps between tiles to reduce these errors.

We have tested the ATLAS star-galaxy separation by comparison with SDSS 
in the overlap areas and we found that 10\% galaxies were misclassified as stars.
We also found that 15\% stars were misclassified as galaxies and that this higher 
rate was due to problems occurring in interchip gaps where only one sub-exposure 
was available. However, the problem is easily cured by using a  more conservative 
star-galaxy separation line particularly at bright magnitudes.

Finally we also presented galaxy and star number counts in a 250deg$^2$
area at the SGP and found where they turned over to define effective
completeness limits. The galaxy counts turn over about at about our
quoted 5$\sigma$ mag limits for galaxies.  Star counts turn over at
about the same limits but this is mainly  dictated by the star-galaxy
separation algorithm.

We conclude that the VST ATLAS is ready for scientific analysis. Currently 
the survey is 75\% complete and it should be 100\% complete by the end of 2015.

%=============================================================%
\section*{ACKNOWLEDGEMENTS}
%=============================================================%

We are grateful to European Southern Observatory (ESO) and their associated
service time VLT Survey Telescope observers for carrying out the  ATLAS
observations. We are indebted to the Cambridge Astronomical Surveys Unit
(CASU) for reducing the ATLAS images and producing catalogues. We also
acknowledge the work of the Wide Field Astronomy Unit (WFAU) at
Edinburgh for archiving the ATLAS data. We also acknowledge use of
publicly available SDSS data. This research has also made use of the
APASS database, located at the AAVSO web site. Funding for APASS has
been provided by the Robert Martin Ayers Sciences Fund.

%---------------------------------------------------------------------

\setlength{\bibhang}{2.0em}
\setlength\labelwidth{0.0em}

\end{document}